\documentclass[twocolumn,preprintnumbers,prb,fleqn]{revtex4}
\usepackage{graphicx,amsfonts,amsbsy}
\usepackage[dvips]{epsfig,color}
\begin{document} 
\title{Spin dynamics in the diluted ferromagnetic Kondo lattice model}
%Disorder and dilution effects in the ferromagnetic Kondo lattice model
\author{Avinash Singh,$^{1,2,3,*}$ Subrat K. Das,$^3$ Anand Sharma,$^{1}$ and Wolfgang Nolting,$^{1}$}
\affiliation{$^1$Institut f\"{u}r Physik, Humboldt-Universit\"{a}t zu Berlin, Newton str. 15, D-12489 Berlin}
\affiliation{$^2$Max-Planck-Institut f\"{u}r Physik Komplexer Systeme, N\"{o}thnitzer str. 38, D-01187 Dresden}
\affiliation{$^3$Department of Physics, Indian Institute of Technology Kanpur - 208016}
\email{avinas@iitk.ac.in}
\begin{abstract}
The interplay of disorder and competing interactions is investigated 
in the carrier-induced ferromagnetic state of the Kondo lattice model
within a numerical finite-size study in which disorder is treated exactly. 
Competition between impurity spin couplings, stability of the ferromagnetic state, 
and magnetic transition temperature are quantitatively investigated 
in terms of magnon properties for different models including dilution, disorder, and weakly-coupled spins. 
A strong optimization is obtained for $T_c$ at hole doping $p << x$, 
highlighting the importance of compensation in diluted magnetic semiconductors.
The estimated $T_c$ is in good agreement with experimental results for $\rm Ga_{1-x}Mn_x As$
for corresponding impurity concentration, hole bandwidth, and compensation.
Finite-temperature spin dynamics is quantitatively studied 
within a locally self-consistent magnon renormalization scheme,
which yields a substantial enhancement in $T_c$ due to spin clustering,
and highlights the nearly-paramagnetic spin dynamics of weakly-coupled spins. 
The large enhancement in density of low-energy magnetic excitations 
due to disorder and competing interactions results in a strong thermal decay of magnetization,
which fits well with the Bloch form $M_0(1-BT^{3/2})$ at low temperature, with $B$ of same order of magnitude 
as obtained in recent squid magnetization measurements on $\rm Ga_{1-x}Mn_x As$ samples.
\end{abstract}
\maketitle
\section{Introduction}
The discovery of ferromagnetism in diluted magnetic semiconductors (DMS) 
such as $\rm Ga_{1-x} Mn_x As$,\cite{ohno2,matsu} 
%and $\rm Ga_{1-x} Mn_x N$, (Ref. 5),
%\cite{Sonoda}
%GaMnAs1,GaMnAs2
%GaMnN3
with transition temperature $T_c \simeq 110$ K for Mn concentration $x \simeq 5\%$,\cite{matsu,ohno3} 
and $\simeq 150$ K in films for $x$ in the range $6.7-8.5\%$,\cite{ku,edmonds} 
has generated tremendous interest not only in view of potential technological applications, but also due to the 
novel ferromagnetism exhibited by these systems in which magnetic interactions between localized spins are mediated by doped carriers.
%\cite{Akai,Dietl1,Taka1,Jung1,Dietl2,Jung2,Dietl3,Das1,Das2,MacD,Konig1,Berciu1,
%Alvarez,dms,Singh2,Timm,Taka2,Pandey,Awsch}
The long-range oscillatory nature of the carrier-mediated spin couplings
results in a variety of interesting behaviour such as significant sensitivity of spin stiffness and 
transition temperature $T_c$ on carrier concentration, 
competing antiferromagnetic interaction and noncollinear ordering,
spin-glass behaviour, spin clustering and disorder-induced localization 
etc.,\cite{mf1,mf2,mf3,sw1,dis2,dmft1,sw2,dis1,dis3,dis6,anomalous,mc3,dassarma,dms,squid}
as recently reviewed.\cite{bhatt,timm}

DMS such as $\rm Ga_{1-x}Mn_x As$ are mixed spin-fermion systems 
in which the $S=5/2$ Mn$^{++}$ impurities replace Ga$^{+++}$, 
thereby contributing a hole to the semiconductor valence band. 
However, large compensation due to As antisite defects reduces the hole density $p$
to nearly $10 \%$ of Mn concentration $x$, which plays a key role in the stabilization 
of long-range ferromagnetic order,
and also provides a complimentary limit to Kondo systems.
The interplay between itinerant carriers in a partially filled band 
and the localized moments is conventionally 
studied within a diluted ferromagnetic Kondo lattice model (FKLM),
wherein $- J {\bf S}_I . {\mbox{\boldmath $\sigma$}}_I $ 
represents the exchange interaction between the localized magnetic impurity spin ${\bf S}_I$ 
and the itinerant electron spin ${\mbox{\boldmath $\sigma$}}_I$.

Recently, finite-temperature spin dynamics due to thermal spin-wave excitations has been studied in 
Ga$_{1-x}$Mn$_x$As samples with different Mn content (thickness about 50nm, Mn content ranging from $2\%$ to $6\%$)
using SQUID (superconducting quantum interference device) magnetization measurements.\cite{sperl} 
The temperature dependence of (low-field) spontaneous magnetization shows nearly linear fall off, 
similar to earlier results exhibiting even a distinct concave behaviour for unannealed samples,\cite{ohno3,ku}
possibly resulting from spin re-orientation transitions due to temperature-dependent magnetic 
anisotropies.\cite{sawicki}
However, the spontaneous magnetization obtained using linear extrapolation from a 0.3-0.4 T magnetic field
to overcome the anisotropy fields, 
discussed earlier for epitaxial ultra-thin Fe and FeCo films,\cite{kipferl}
is found to be well described by the Bloch form $M(T)=M_0(1-BT^{3/2})$, 
with a spin-wave parameter $B \sim 1-3\times10^{-3}\; {\rm K}^{-2/3}$ 
which is about two orders of magnitude higher than for Fe and FeCo films. 
This large difference cannot be attributed only to a reduced exchange interaction. 
Post-growth annealing has been shown to significantly increase $T_c$, possibly due to enhancement of carrier concentration
resulting from decrease of Mn interstitial concentration.\cite{ku} 
For a 50nm sample with $6\%$ Mn,
a decrease in $B$ from $2.7\times10^{-3}\; {\rm K}^{-2/3}$ to $1.4\times10^{-3}\; {\rm K}^{-2/3}$ 
has also been obtained upon annealing,\cite{sperl} 
with a corresponding increase in the spin-wave stiffness constant $D$
from 53 meV$\AA ^2$ to 71 meV$\AA ^2$,
which is of same order of magnitude as obtained from magnetic Kerr measurements using pump-probe setup
of standing spin waves in ferromagnetic $\rm Ga_{1-x}Mn_x As$ thin films.\cite{notredame} 

In view of these recent findings of strong thermal decay of magnetization in DMS systems, 
it is of interest to investigate the interplay of disorder and competing interactions on magnon excitations which typically constitute the lowest-energy excitations 
and therefore essentially determine the low-temperature behaviour in magnetic materials.   
In this paper we consider the ferromagnetic Kondo lattice model (FKLM) with several extensions;
we treat disorder exactly by considering explicit disorder realizations on finite-size systems
(random impurity locations, random on-site potential etc.) 
and average over sufficiently large number of configurations to obtain statistically reliable results. 
We also quantitatively investigate finite-temperature spin dynamics due to thermal excitation of magnons
within a locally self-consistent magnon renormalization scheme, 
equivalent to a site-dependent Tyablikov decoupling (local RPA),
and present the first site-dependent calculations for local impurity magnetization 
by explicitly incorporating the spatial feature of magnon states. 
As we shall see, the disorder-induced formation of low- and high-energy localized magnon modes, 
corresponding to weakly- and strongly-coupled spins respectively, 
results in a variety of interesting spin-dynamics behaviour, 
such as concave magnetization behaviour due to dominant nearly-paramagnetic contribution of weakly-coupled spins 
and an enhancement in $T_c$ due to strong local correlations in impurity-spin clusters.
A simplified calculation involving two discrete spin-excitation energy scales and a finite fraction 
of weakly-coupled spins was discussed earlier.\cite{squid} 

Magnon excitations provide a composite measure of the carrier-induced
spin couplings in the collinear ferromagnetic state, with negative-energy modes signalling
instability due to competing antiferromagnetic (AF) spin interactions.
Magnon properties have been studied as function of electron density $n$ in the conduction band 
and the spin-fermion coupling $J$ within the concentrated FKLM (having a magnetic impurity at every lattice site)
in the context of heavy fermion materials,\cite{sig} ferromagnetic metals Gd, Tb, Dy, 
doped EuX\cite{donath} and manganites.\cite{furu,wang,yunoki,vogt} 
In the context of DMS, magnon properties have been studied earlier in the random phase approximation (RPA) 
for the impurity-band model\cite{dis2} and for the diluted Hubbard model\cite{dms,squid} 
where disorder was treated exactly within finite-size numerical studies,
and for the diluted FKLM within the virtual crystal approximation (VCA) 
where a uniform impurity-induced Zeeman splitting of the carrier spin bands is assumed,\cite{sw1}
within the coherent potential approximation (CPA),\cite{sw2,nol2}
and also for ordered impurity arrangements to make quantitative comparisons with different approximations.\cite{diluted}
Magnon spectrum and transition temperature have also been obtained recently 
for $\rm Ga_{1-x} Mn_x As$ and $\rm Ga_{1-x} Mn_x N$
in terms of effective Heisenberg models with realistic exchange couplings,
obtained recently from first-principle calculations as well.\cite{hilbert_nolting,bouzerar}

Within mean-field (MF) approaches for the magnetization behaviour in DMS,\cite{dis2,anomalous,dassarma}
it is the large separation of spin and fermion energy scales which plays the key role.
For hole concentration $p$ in the concentrated FKLM,
the very low effective ordering field for spins $(Jp)$ compared to that for fermions $(JS)$ 
results in an intermediate-temperature regime  $Jp \ll k_{\rm B}T \ll JS $ 
wherein the nearly-paramagnetic contribution of spins
accounts for the unusual concave shape.\cite{dis2,anomalous,dassarma} 
%In the diluted case, holes tend to accumulate near impurity spins,
%resulting in enhanced local fermion polarization $\sim p/x$ per impurity,
%which narrows this intermediate-temperature regime to $Jp/x \ll k_{\rm B}T \ll JS $.
Every impurity spin contributes equally in this picture and disorder has no direct role,
as in other VCA-based magnetization studies within spin-wave theory.\cite{sw1,sw2}
However, the magnon energy scale is typically an order of magnitude smaller than the MF low-energy scale $Jp$, 
even in the disordered case where fermion spin polarization is substantially reduced in impurity-poor regions, 
implying that the relevant energy/temperature scales for spin dynamics are actually determined by these collective excitations. 
Unlike the purely local mean field $Jp$, the collective excitations involve non-local spin couplings $J_{ij}$, 
and hence not only incorporate competing interactions and frustration effects, 
but also differentiate between strongly-coupled cluster spins and weakly-coupled isolated spins. 
As only the finite fraction of weakly-coupled spins involved in the low-energy localized magnon states 
yield the nearly paramagnetic contribution, this picture is quite different from the MFT picture.

The organization of this paper is as follows. The RPA-level theory for magnon excitations in real space 
is derived in section II for a general fermion Hamiltonian, including dilution, disorder, and multiple bands,
and the Goldstone-mode behaviour expected from spin-rotation symmetry is explicitly verified.
Results for several different ferromagnetic Kondo lattice models
--- involving dilution, disorder, and weakly-coupled spins ---
are then discussed in Sections III, V, and VI, respectively,
with finite-temperature spin dynamics introduced in section IV.
Conclusions are presented in section VII. 

\section{Magnon excitations}
Magnons represent transverse spin fluctuations about the spontaneously broken-symmetry state
and constitute gapless, low-energy excitations for magnetic systems possessing continuous spin-rotational symmetry.
At low temperature, magnons therefore play an important role in diverse macroscopic 
properties such as existence of long-range order, magnitude and temperature dependence 
of the order parameter, magnetic transition temperature, spin correlations etc. 
In the following we consider finite temperature $T$,
and obtain magnon excitations at the RPA level where magnons interactions are neglected.
Finite impurity concentration, disorder, and interaction strength are treated on an equal footing in this approach.

We consider the Kondo lattice model 
\begin{equation}
H = H_0 - \frac{J}{2} \sum_I{\bf S}_I.{\mbox{\boldmath $\sigma$}}_I
\end{equation}
where $H_0$ represents the free-fermion part consisting generally of hopping and on-site energy terms,
and the second term represents the exchange coupling between impurity spins ${\bf S}_I$ and fermion spins ${\mbox{\boldmath $\sigma$}}_I /2$
at impurity sites $I$. The analysis for magnon excitations discussed below is independent of details of $H_0$, 
for which several different cases including hopping and potential disorder are considered in later sections.

Applying the approximate Holstein-Primakoff transformation 
from the spin-lowering ($S_I ^-$) and spin-raising ($S_I^+$) operators 
to boson (magnon) creation and annihilation operators $b_I^\dagger$ and $b_I$,
\begin{eqnarray}
S_I ^+ &=&  b_I \sqrt{2S_I}  \nonumber \\
S_I ^- &=&  b_I ^\dagger \sqrt{2S_I} \nonumber \\
S_I ^z &=& S_I - b_I ^\dagger b_I 
\end{eqnarray}
the Kondo lattice Hamiltonian reduces to  
\begin{eqnarray}
& & H = H_0 \nonumber \\
& & -\frac{J}{2} \sum_I \left [ \frac{\sqrt{2S_I}}{2} 
\left (b_I\sigma_I^- + b_I^\dagger \sigma_I^+ \right )
+ \left (S_I -  b_I^\dagger b_I \right ) \sigma_I^z \right ] \; ,
\end{eqnarray}
where $\sigma_I^\pm \equiv \sigma_I ^x \pm i \sigma_I ^y$
and  the "spin quantum numbers" $S_I \equiv \langle S_I ^z \rangle_{\rm MF}$
refer to finite-temperature magnetizations obtained self consistently in the mean-field state.
The above approximate transformation neglects quartic magnon interaction terms of order $1/S$.

Starting with a MF approximation,
$\langle b_I^\dagger \rangle = \langle b_I \rangle = 
\langle b_I^\dagger b_I \rangle = 0$, 
the Hamiltonian (3) decouples into a fermion part with an impurity-field term
\begin{equation}
{\cal H}^0 _{\rm fermion} = H_0 -  \frac{J}{2} \sum_I S_I \sigma_I^z 
\end{equation}
and a local boson part  
\begin{equation}
{\cal H}^0 _{\rm boson} = \frac{J}{2}\sum_I \langle \sigma_I^z \rangle  b_I^\dagger b_I 
\equiv \sum_I {\cal E}_I \; b_I^\dagger b_I \; ,
\end{equation}
representing the energy cost of a local spin deviation.
At the MF level, determination of impurity and fermion magnetizations
$\langle S_I ^z \rangle$ and $\langle \sigma_I^z \rangle$
involves a self-consistent solution of the coupled spin-fermion problem 
in terms of Brillouin and Fermi functions.
In the zero-temperature limit, as $S_I \equiv \langle S_I ^z \rangle_{\rm MF} = S$,
the impurity magnetic field seen by fermions has same magnitude $JS/2$ on all sites,
however, the magnetic field  $J\langle \sigma_I^z \rangle_{\rm MF}/2$ seen by impurity spins remains non-uniform 
due to positional disorder. 

Proceeding next to transverse spin fluctuations about the MF state,
we obtain the time-ordered magnon propagator for the impurity spins
in terms of the corresponding boson propagator 
\begin{eqnarray}
& & {\cal G}_{IJ}^{+-}(t-t') = i\langle \Psi_{\rm G} \mid T[S_I ^+ (t) S_J ^- (t')]\mid 
\Psi_{\rm G} \rangle \nonumber \\
& & = \sqrt{2S_I} \left ( i\langle \Psi_{\rm G} \mid T[b_I (t) b_J ^\dagger (t')]\mid 
\Psi_{\rm G} \rangle \right ) \sqrt{2S_J} 
\end{eqnarray}
at the RPA level by summing over all bubble diagrams 
%%%%%%%%%%%%%%%%%%%%%%%%%%%%%%%%%%%%%%%%%%%%%%%%%%%%%%
\begin{figure}[hbt]
\input{chipm1.pstex_t}
\end{figure}
%%%%%%%%%%%%%%%%%%%%%%%%%%%%%%%%%%%%%%%%%%%%%%%%%%%%%%
\ \\
%\hspace*{-5mm} 
where the particle-hole bubble
\begin{eqnarray}
[\chi^0(\omega)]_{IJ} &=& i \int \frac{d\omega'}{2\pi}[G^{\uparrow}(\omega')]_{IJ}[G^{\downarrow}(\omega'-\omega)]_{JI} \nonumber \\ 
&=& \sum_{l,m}
\frac{\psi_{l\uparrow}^I \psi_{l\uparrow}^J \psi_{m\downarrow}^I \psi_{m\downarrow}^J }
{E_{m\downarrow} - E_{l\uparrow} + \omega} \; f_{l\uparrow}(1-f_{m\downarrow}) \nonumber \\
&+&
\sum_{l,m}
\frac{\psi_{l\uparrow}^I \psi_{l\uparrow}^J \psi_{m\downarrow}^I \psi_{m\downarrow}^J }
{E_{l\uparrow} - E_{m\downarrow} - \omega} \; (1-f_{l\uparrow})f_{m\downarrow}  
\end{eqnarray}
with Fermi functions $f_{l\uparrow}$ and $f_{m\downarrow}$ 
involves integrating out the fermions (eigenvalues $\{E_{l\sigma}\}$ and wave functions $\{\psi_{l\sigma}\}$)
in the broken-symmetry state.
It is the particle-hole bubble $[\chi^0(\omega)]_{IJ}$ 
which mediates the carrier-induced impurity spin couplings in the ferromagnetic state, 
and the oscillatory, long-range nature of the spin couplings is 
effectively controlled by the fermion band filling and the impurity field strength.

\begin{figure}
\hspace*{-3mm}
\includegraphics[width=90mm]{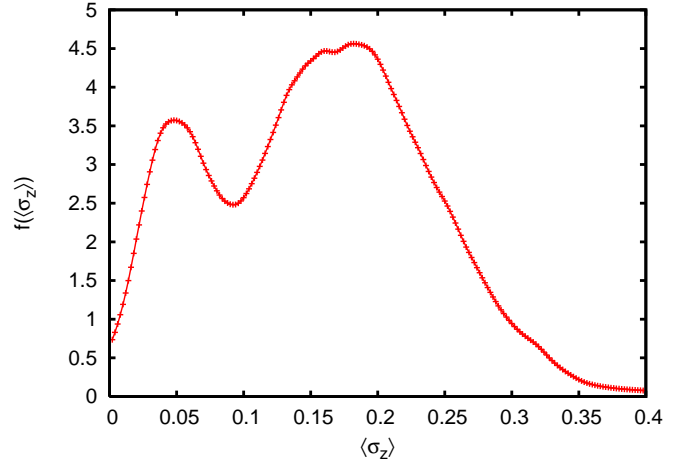}
\caption{Distribution of fermion spin densities on impurity sites, 
showing significantly reduced densities in impurity-poor regions relative to their average value $\approx p/x$. 
Here $J=4$, hole doping $p \approx 4\%$ and impurity concentration $x \approx 20\%$.}
\end{figure}

In terms of the site-diagonal zeroth-order magnon propagator 
\begin{equation}
[{\cal G}^0(\omega)]=\frac{[2S_I]}{\omega - {\cal H}^0 _{\rm boson}} 
= \sum_I \frac{2S_I}{\omega - {\cal E}^0 _I + i\eta} |I\rangle \langle I| 
\end{equation}
the full magnon propagator can then be expressed as
\begin{eqnarray}
[{\cal G}^{+-}(\omega)] &=& 
\frac{[{\cal G}^0(\omega)]}
{1 + \frac{J^2}{4} [\chi^0(\omega)] [{\cal G}^0(\omega)] } \nonumber \\
&=& [\sqrt{2S_I}] \left ( \frac{1}{\omega - [{\cal H}(\omega)]} \right )[\sqrt{2S_J}] 
%= \frac{[2S_I]}{\omega - [{\cal H}(\omega)]} 
\end{eqnarray}
in terms of a boson "Hamiltonian" 
\begin{equation}
[{\cal H}(\omega)]_{IJ} =  {\cal E}_I \delta_{IJ} - {\cal E}_{IJ}(\omega)
\end{equation}
involving the boson on-site energy
\begin{equation}
{\cal E}_I \equiv \frac{J}{2} \langle \sigma_I^z \rangle 
\end{equation}
and the boson hopping terms 
\begin{equation}
{\cal E}_{IJ}(\omega) \equiv \sqrt{2S_I} \left (\frac{J^2}{4}  [\chi^0(\omega)]_{IJ} \right ) \sqrt{2S_J}
\end{equation}
associated with carrier-induced spin couplings ${\cal J}_{IJ} \sim (J^2/4) [\chi^0(\omega)]_{IJ}$.
While dynamical effects are in principle included in the magnon Hamiltonian $[{\cal H}(\omega)]$,
we find that the $\omega$ dependence is sufficiently weak to be neglected, so that
the eigenvalues and eigenvectors of $[{\cal H}]$ directly yield 
the (bare) magnon energies $\{\omega_l^0\}$ and wave functions $\{\phi_l^0\}$.

Equation (9) has exactly same structure as obtained from the Tyablikov decoupling for an effective Heisenberg 
model with spin couplings ${\cal J}_{IJ} \sim (J^2/4) [\chi^0(\omega)]_{IJ}$, 
and readily yields a locally self-consistent renormalized magnon theory, as discussed in section IV. 

In order to obtain a zero-energy Goldstone mode (boson amplitude $\sqrt{2S_I}$) consistent with spin-rotation symmetry, 
the energy cost of creating a local spin deviation 
must be exactly offset by the delocalization-induced energy gain,
which requires that 
\begin{equation}
{\cal E}_I = \sum_J \frac{J^2}{4} [\chi^0 (\omega=0)]_{IJ} . 2S_J\; ,
\end{equation}
which is indeed exactly satisfied. 
This is easily verified in the concentrated limit where translational symmetry results in 
plane-wave fermion states with band energies 
$E_{{\bf k}\sigma}= \epsilon_{\bf k} - \sigma J\langle S^z \rangle /2$, 
and the particle-hole propagator (7) simplifies to 
\begin{eqnarray}
\chi^0(q,\omega=0) &=& \sum_J [\chi^0(\omega=0)]_{IJ} \nonumber \\
&=& \sum_{\bf k} 
\frac{f_{{\bf k}\uparrow} (1-f_{{\bf k}\downarrow})} {J\langle S_z\rangle}
+ 
\frac{(1-f_{{\bf k}\uparrow}) f_{{\bf k}\downarrow}} {-J\langle S_z\rangle} \nonumber \\
&=& \sum_{\bf k} 
\frac{(f_{{\bf k}\uparrow} - f_{{\bf k}\downarrow}) }
{J\langle S^z\rangle} = \frac{\langle \sigma^z \rangle}{J\langle S^z\rangle}
\end{eqnarray}
which ensures that the required condition (13) is satisfied.
The required condition (13) for a spin-rotationally-invariant system can be easily derived from a perturbation 
analysis for the transverse fermion-spin density due to small transverse impurity fields
(corresponding to small twist of the spin coordinate system) and then comparing with that expected on 
symmetry grounds. 

The transverse spin-fluctuation propagator for the fermion spin,
defined in analogy with (6) in terms of fermion spin operators $\sigma^-$ and $\sigma^+$, 
involves the same bubble sum at the RPA level and is given by  
\begin{equation}
[\chi^{+-}(\omega)] = \frac{[\chi^0(\omega)]}
{1 + \frac{J^2}{4} [{\cal G}^0(\omega)][\chi^0(\omega)]} \; .
\end{equation}
While the fermion spin-fluctuation propagator also involves single-particle (Stoner) excitations, 
described by the poles of $[\chi^0(\omega)]$,
the collective (magnon) excitations are, as expected, same as for the impurity spin-fluctuation propagator,
as it is the fermions which mediate the impurity spin couplings.  
We note that the above fermion propagator involves an effective local (but dynamical) fermion interaction 
\begin{equation}
[{\cal U}(\omega)] = \frac{J^2}{4} [{\cal G}^0(\omega)] 
\end{equation}
corresponding to the local magnon exchange.
Furthermore, an effective isotropic fermion interaction 
${\cal U}(\omega) {\mbox{\boldmath $\sigma$}}_I . {\mbox{\boldmath $\sigma$}}_I /4$, 
expected from spin-rotational invariance in the fermion sector,
provides not only the required transverse terms (15,16) 
but also the correct HF description (4) from the longitudinal term $\sigma_I ^z \sigma_I ^z$.
These observations allow for a spin-rotationally-symmetric analysis of quantum corrections 
to the magnon propagator due to self-energy and vertex corrections 
along same lines as studied recently within the Hubbard model.\cite{vertex}
 
For a multiband case, with an exchange coupling 
$- \frac{1}{2} \sum_{I\alpha} J_\alpha {\bf S}_I.{\mbox{\boldmath $\sigma$}}_{I \alpha}$ 
involving sum over the fermion band index $\alpha$, 
a straightforward generalization of the above analysis yields the same structure (9) for the magnon propagator,
with magnon on-site and hopping energy terms given by
\begin{equation}
{\cal E}_I = \frac{1}{2} \sum_\alpha J_\alpha \langle \sigma_I^z \rangle_\alpha 
\end{equation}
\begin{equation}
{\cal E}_{IJ}(\omega)=\frac{1}{4} \sum_{\alpha\beta} J_\alpha J_\beta [\chi^0(\omega)]_{IJ} ^{\alpha\beta}  
\; . \sqrt{2S_I} \sqrt{2S_J}
\end{equation}
If the free-fermion Hamiltonian $H_0$ does not involve any band mixing, 
as considered in recent studies,\cite{dag06}
then $ [\chi^0(\omega)]_{IJ} ^{\alpha\beta}$ is diagonal in band indices,
and $ [\chi^0(\omega)]_{IJ} ^\alpha$ for band $\alpha$ is simply given by Eq. (7) 
in terms of states for the corresponding fermion band. 
A recent equation-of-motion analysis for the multiband case yields a similar structure for the mRKKY spin couplings.\cite{sharma_06}

The dimension of the magnon Hamiltonian ${\cal H}$ is $N_m$, the number of magnetic impurities per configuration. 
From the $N_m$ magnon energies $\{\omega_l^0\}$ and wave functions $\{\phi_l^0\}$
we evaluate the magnon density of states and participtation ratio (PR) 
\begin{eqnarray}
N(\omega) &=& \frac{1}{\pi} \frac{1}{N_m} 
\sum_l \frac{\eta}{(\omega-\omega_l^0)^2 + \eta^2} 
\;\;\;(\eta\rightarrow 0) \nonumber \\
{\rm PR} &=& 1/\sum_I (\phi_l ^{0I} )^4 \; ,
\end{eqnarray}
which together provide a complete picture of both spectral and spatial features of magnon states. 
The participation ratio provides a measure of the number of impurity spins over which 
the magnon state is extended. 
Generally, if the normalized wave function $\phi_l$ corresponds to a state
with essentially non-zero amplitude $\phi_l^I \sim 1/\sqrt{n}$ on $n$ sites, then PR$\sim n$. 
The participation ratio for magnon states can thus range between $N_m$ for a fully extended magnon state 
and 1 for a site-localized magnon state. 
The PR thus readily allows localized magnon states (PR$\sim 1$)  
to be distinguished from extended magnon states (PR$\sim N_m$). 

In order to quantitatively study the interplay of disorder and competing interactions,
and the role of disorder on spin dynamics
we consider below different disorder models for the fermion Hamiltonian $H_0$ 
on finite-size lattices with explicit disorder realizations of impurity position, on-site energy etc.
so that disorder effects are treated exactly here. 
We consider throughout a simple cubic lattice for simplicity, with periodic boundary conditions.
The number of host lattice sites $N=L^3$ determines the dimension of the fermion Hamiltonian 
$H_0$ to be diagonalized, and we have considered system sizes $L=8,10,12$.
In all cases we consider the saturated ferromagnetic state with number of spin-$\uparrow$ fermions 
$N_\uparrow = N$ and doping in the spin-$\downarrow$ band
which is pushed up by the impurity Zeeman field. 
The impurity and carrier (hole) concentrations referred below 
correspond to $x=N_m/N$ and $p=(N-N_\downarrow)/N$, respectively.

\section{Diluted Kondo Lattice Model}
Providing a minimal description of the exchange coupling in DMS systems 
such as $\rm Ga_{1-x}Mn_x As$ between Mn impurity and carrier spins,
the diluted Kondo lattice model 
\begin{equation}
H = t \sum_{i,\delta,\sigma} a_{i,\sigma}^\dagger a_{i+\delta,\sigma} 
+ \epsilon_d \sum_{I,\sigma}  a_{I,\sigma}^\dagger a_{I,\sigma} 
- \frac{J}{2} \sum_I{\bf S}_I.{\mbox{\boldmath $\sigma$}}_I
\end{equation}
represents $N_m$ magnetic impurities placed randomly on a fraction ($I$) of the $N$ host sites ($i$),
with impurity concentration $x=N_m/N$. 
We consider a positive nearest-neighbour hopping $t$
so that the host ${\bf k}=0$ state lies at the top of the valence band;
doped carriers (holes) go in long wavelength states, 
so that small-$k$ particle-hole processes near the Fermi energy are dominant
in the carrier-induced ferromagnetic spin couplings,
and therefore other details of the energy band are relatively unimportant. 
In the following we set $t=1$ as the unit of energy scale. 
Also, as the temperature regime of interest, set by the magnon energy scale,
is very low compared to the MF energy scale,
in the following we have only considered the $T=0$ case which provides
a good approximation of the low-temperature MF state.

\begin{figure}
\hspace*{-3mm}
\includegraphics[width=90mm]{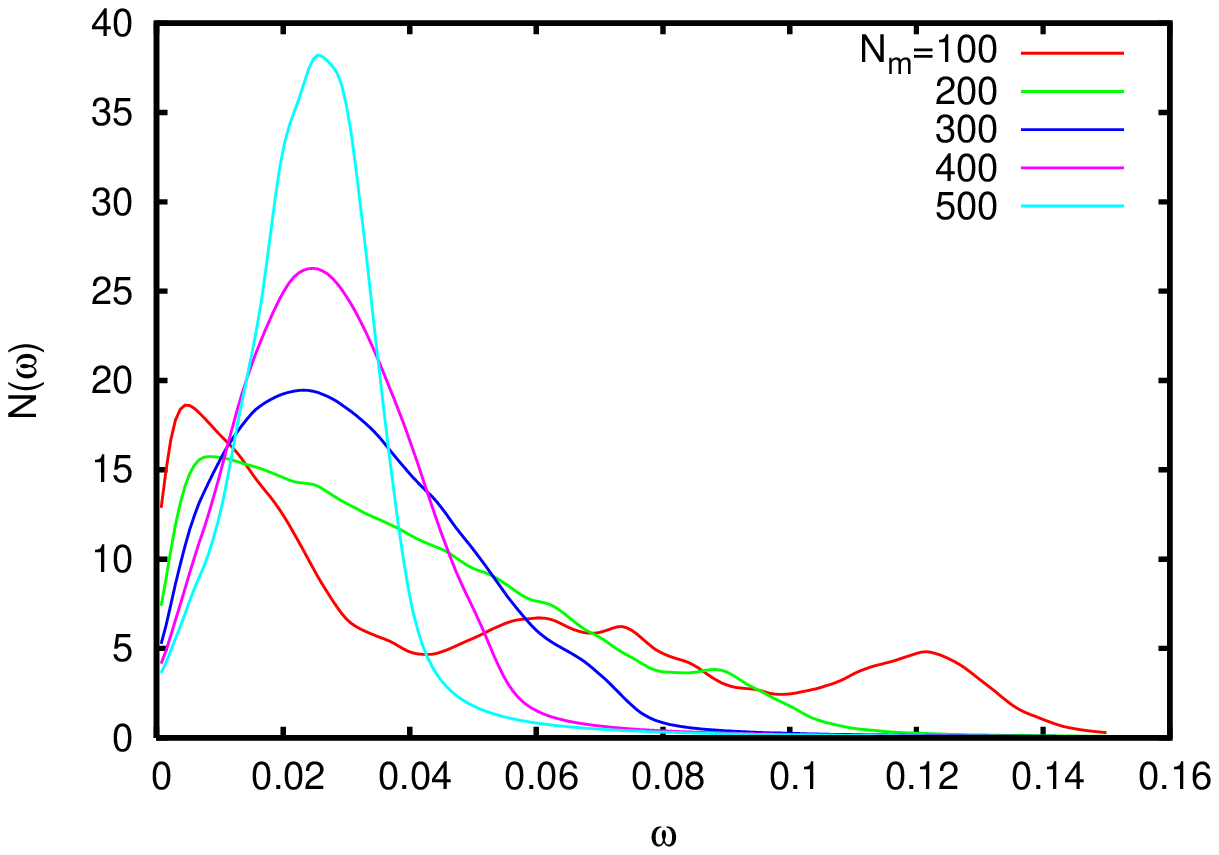}
\hspace*{-3mm}
\includegraphics[width=90mm]{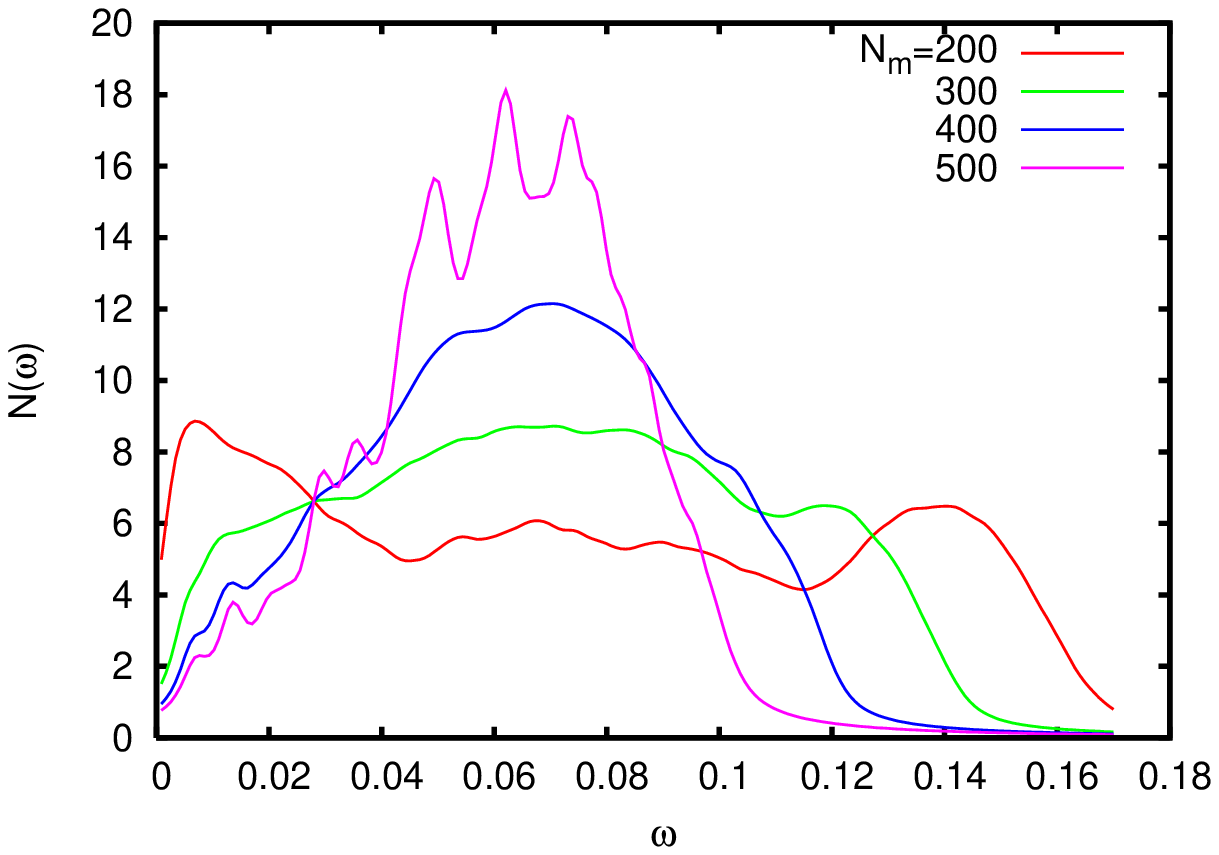}
\caption{Magnon density of states for different impurity dilutions for a $N=8^3$ system with $J=4$.
The upper and lower panels correspond to hole doping $p \approx 4\%$ and $\approx 11\%$, respectively. 
Inset shows (for $x\approx 20\%$) comparison of magnon DOS for different system sizes $N=L^3$.}
\end{figure}
\begin{figure}
\vspace*{-145mm}
\hspace*{10mm}
\includegraphics[width=35mm]{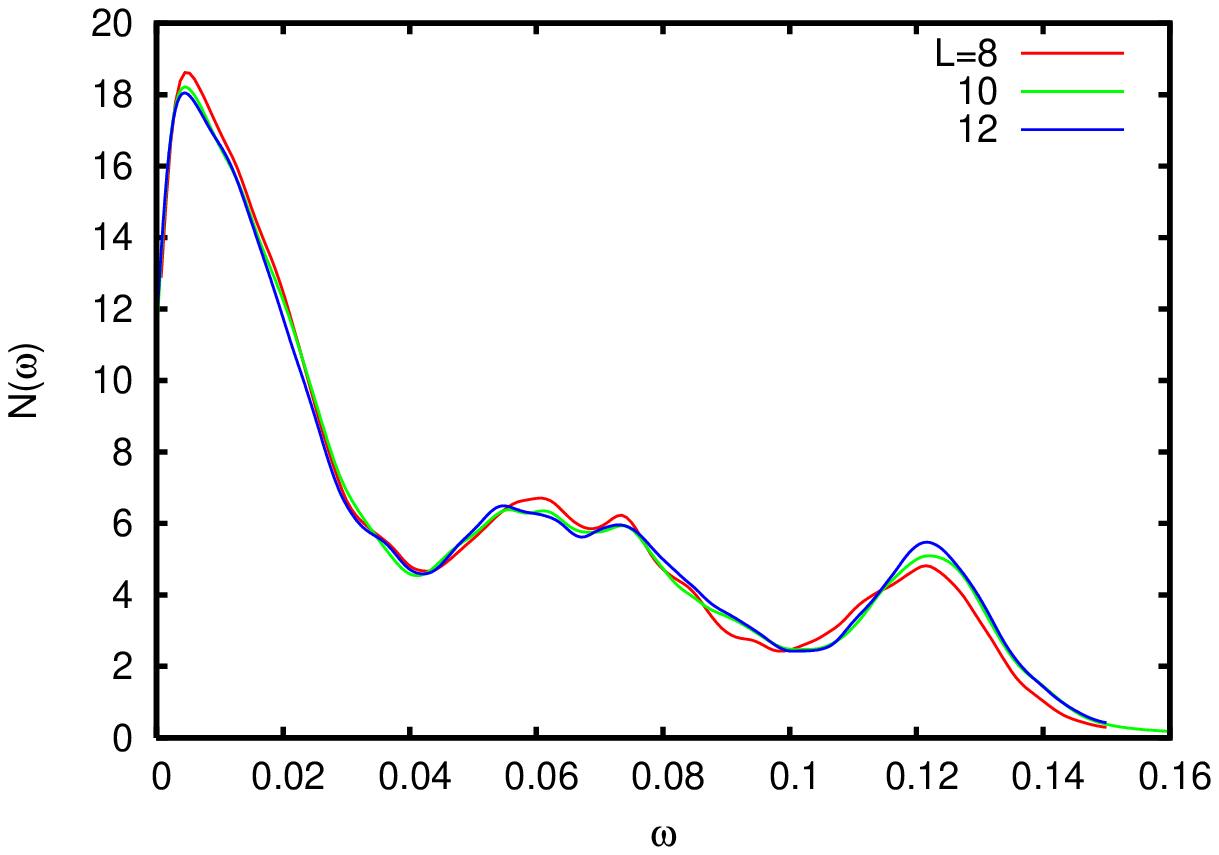}
\vspace*{120mm}
\end{figure}

Fermions see an effective potential disorder due to dilution;
both disorder and dilution cannot be changed independently.
We have therefore included an impurity on-site energy (chemical potential) $\epsilon_d$, 
which 
%in addition to possibly being significant for DMS systems, 
provides an effective control of disorder {\em independently} of dilution.
For negative $\epsilon_d$, the reduced impurity-site mean-field energy 
$(\epsilon_d + JS/2)$ for spin-$\downarrow$ fermions 
reduces the disparity between host and impurity sites, resulting in lower effective disorder 
(for $\epsilon_d = -JS/2$, there is no disorder at MF level!).
The effects are quite dramatic on the magnon spectrum, showing significant decrease in the
density of low-energy modes and hence enhanced stability of the carrier-mediated ferromagnetic state.

The fermion spin polarization $\langle \sigma_I^z \rangle$ typically shows significant site variation 
as fermions tend to accumulate in impurity-rich regions, leaving significantly reduced 
polarization in impurity-poor regions. 
Figure 1 shows the distribution of fermion spin polarization on impurity sites, 
obtained by diagonalizing the fermion Hamiltonian (4) on a $N=8^3$ system for 50 configurations.
Besides the broad peak at the expected value of $p/x$ corresponding to average hole density per impurity site,
there is an additional peak at significantly reduced fermion spin polarization corresponding to 
impurity-poor regions; this introduces a new low-energy MF scale $J\langle \sigma_I^z \rangle S/2$ 
such that at comparable or higher temperatures these weakly bound impurity spins become nearly paramagnetic. 
However, as we shall see, the magnon energy scale is an order of magnitude smaller than even this low-energy MF
scale, indicating that the dominant spin dynamics is due to thermal excitation of the collective magnetic excitations rather than that described by the Brillouin function corresponding to single-spin excitation energies 
within the MF theory. As the temperature regime of interest is therefore much lower than this MF energy scale,
a self-consistent finite-temperature mean-field analysis is not required
and therefore the $T=0$ description provides a good approximation of the low-temperature MF state. 

\begin{figure}
\hspace*{-3mm}
\includegraphics[width=90mm]{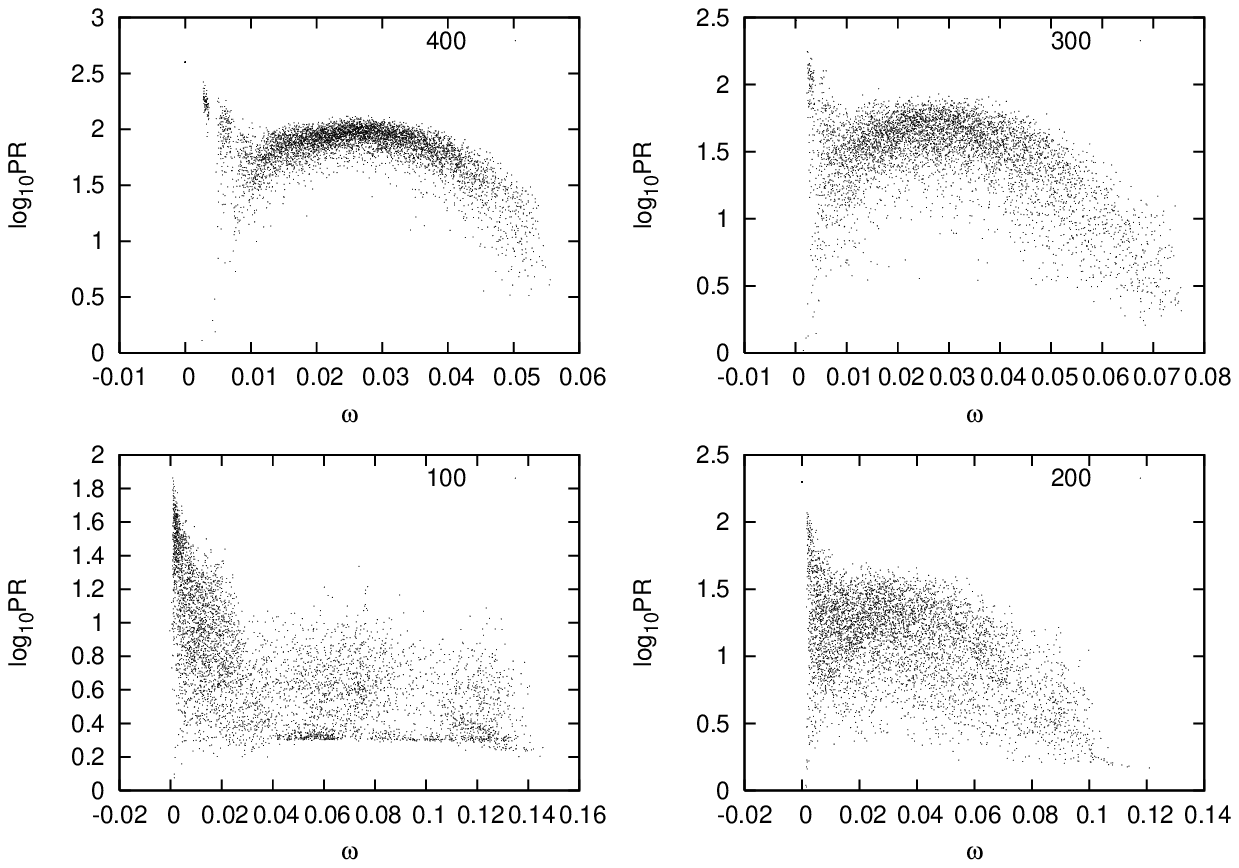}
%\end{figure}
%\begin{figure}
\hspace*{-3mm}
\includegraphics[width=90mm]{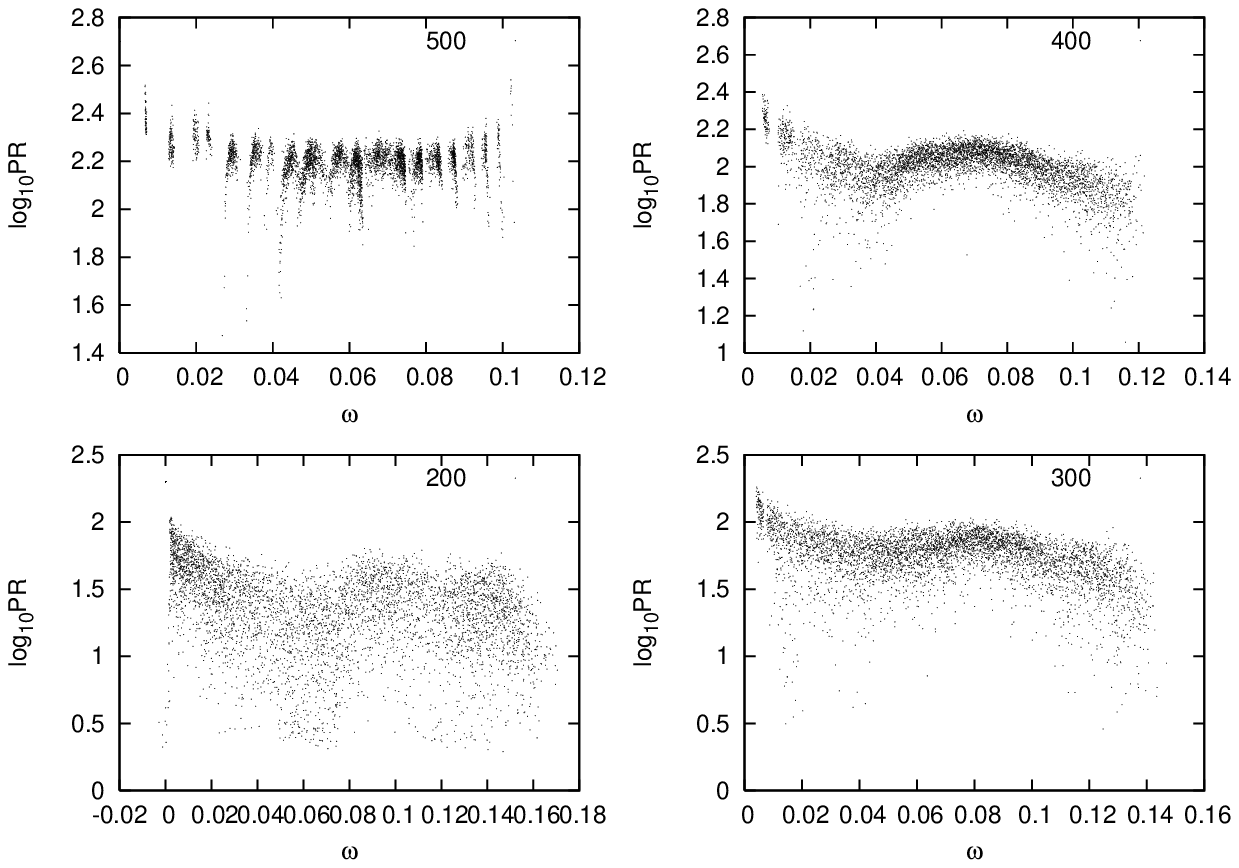}
\caption{Participation ratio for different impurity dilutions for a $N=8^3$ system with $J=4$
and hole concentrations $p \approx 4\%$ and $\approx 11\%$
in upper (four) and lower (four) panels, respectively.}
\end{figure}
%$N_\downarrow = 493$ 
%$N_\downarrow = 455$

Figure 2 shows the configuration-averaged magnon density of states (DOS) 
for two different hole doping concentrations and for varying degree of dilution.
Initially dilution is seen to essentially broaden the magnon spectrum, 
but higher dilution results in significant softening of the collective excitations, 
with the magnon DOS peak progressively shifting to lower energy.
Furthermore, this magnon softening is distinctly more pronounced at lower hole doping,
reflecting more effective competition due to longer range of spin couplings induced at lower hole doping.
In addition, new structure appears at higher energy, particularly at higher dilution, 
which is due to localized magnon states associated with strongly-coupled spins in spin clusters,
as also reported in earlier studies.\cite{dis2,squid}
At still higher dilutions (not shown), negative-energy states appear in the magnon spectrum,
indicating instability of the collinear ferromagnetic state.
The correlation between spectral and spatial features of the magnon states is contained in the PR plots shown in Figure 3
for the two doping levels. In addition to the magnon softening with increasing dilution, 
the PR plots also show the increasing localization of magnon states with dilution,
especially for the high-energy modes which are more likely to be localized in the strongly-coupled spin clusters.

\begin{figure}
\hspace*{-3mm}
\includegraphics[width=90mm]{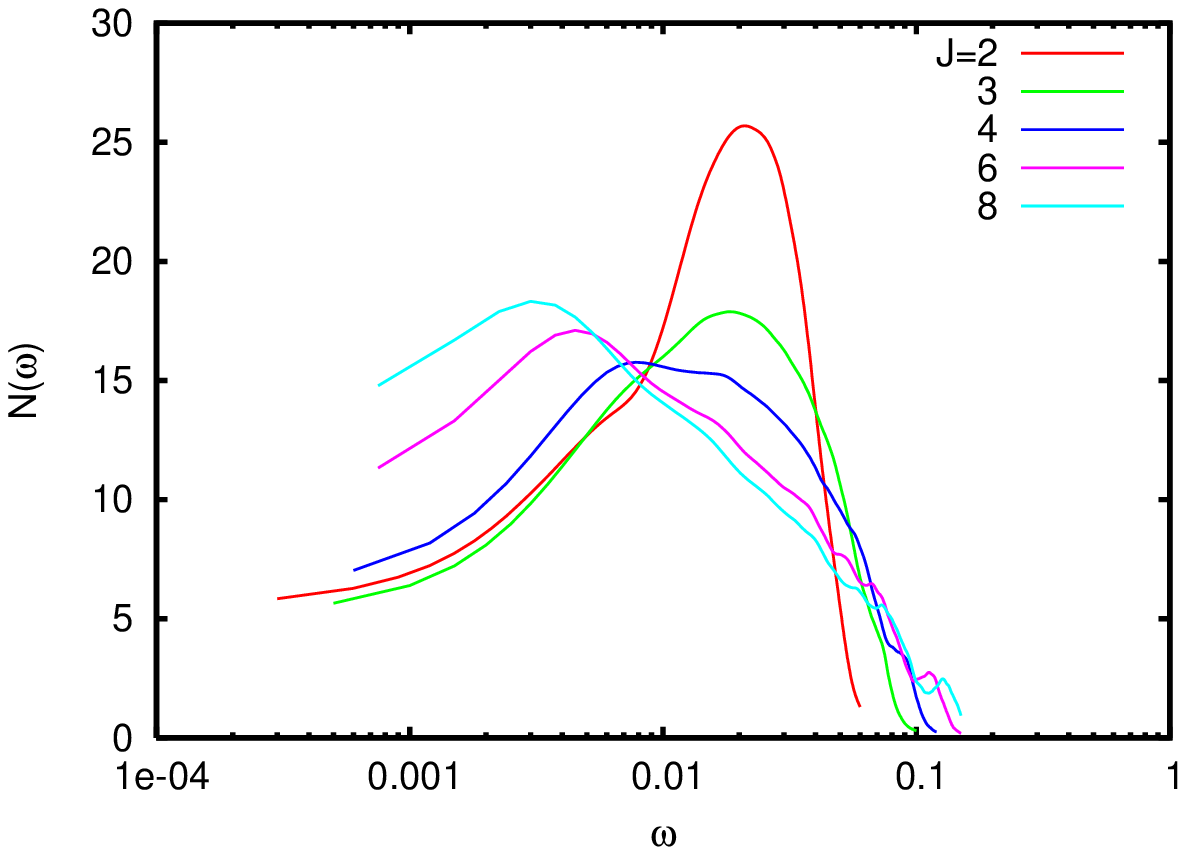}
\hspace*{-3mm}
\includegraphics[width=90mm]{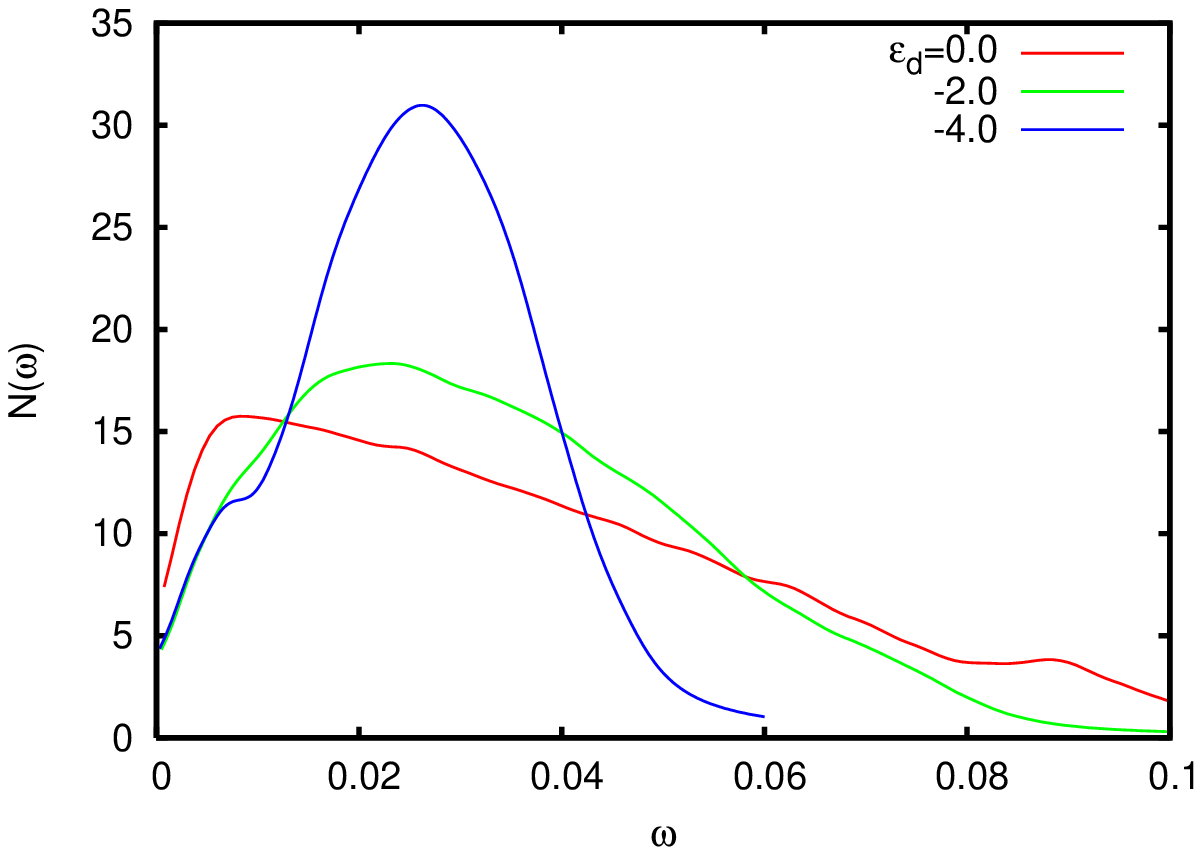}
\caption{Magnon density of states for different $J$ and $\epsilon_d$,
showing effective control of impurity disorder independently of dilution,
for a $N=8^3$ system with $p\approx 4\%$, $x\approx 40\%$, 
and $\epsilon_d = 0$ and $J=4.0$ in upper and lower panels, respectively.}
\end{figure}

We note here that the magnon energy scale for extended modes, 
which essentially determines the spin dynamics in the ordered state due to thermal excitations of magnons, 
is very low compared to the hopping energy scale,
as low as around $\sim 0.02$ for the $x\approx 20\%$ case in Fig. 2.
On the other hand, corresponding to reduced fermion polarization in impurity-poor regions,
the low-energy MF scale is $J\langle \sigma_I^z \rangle S/2 \sim 0.25$. 
The nearly order-of-magnitude separation between these two energy scales 
implies, as discussed above, that the $T=0$ calculations provide a good
description of the low-temperature MF state.

Figure 4 shows effective control of impurity disorder at fixed dilution and doping.
Enhanced potential disorder with increasing $J$ results in magnon softening (upper panel), 
whereas negative $\epsilon_d$ effectively reduces disorder in the doped spin-$\downarrow$ band 
and results in magnon stiffening (lower panel), 
indicating stabilization of the ferromagnetic state and higher $T_c$. 

Figure 5 shows the strong optimization of the transition temperature $T_c$ with carrier concentration due to the
characteristic competition between increasing overall magnitude of the carrier-induced spin couplings
$J^2 \chi^0 _{IJ} $ and the increasing rapidity of its oscillation. 
Here we have estimated $T_c$ from the configuration average
\begin{equation}
\frac{1}{T_c} = \left < \frac{1}{N_m}\sum_l \frac{1}{\omega_l^0} \right >_c
\end{equation}
and taken a hole bandwidth ($W=12t = 10$ eV) of the order of that of GaAs.
Also, we have taken $x=20\%$ for the impurity concentration (per unit cell for our sc lattice), 
corresponding to $5\%$ Mn in the fcc system $\rm Ga_{1-x} Mn_x As$,
which has 4 Ga sites per unit cell, and therefore impurity concentration $4x$ per unit cell.
The peak at $p/x \sim 1/5$ shows the importance of compensation in DMS systems,
and the calculated $T_c$ is in good agreement 
with experimental results for $\rm Ga_{1-x} Mn_x As$. 

%The potential disorder increases with dilution and $J$, and the fermion states become increasingly localized near spin %clusters and isolated spins. 

\begin{figure}
\hspace*{-3mm}
\includegraphics[width=90mm]{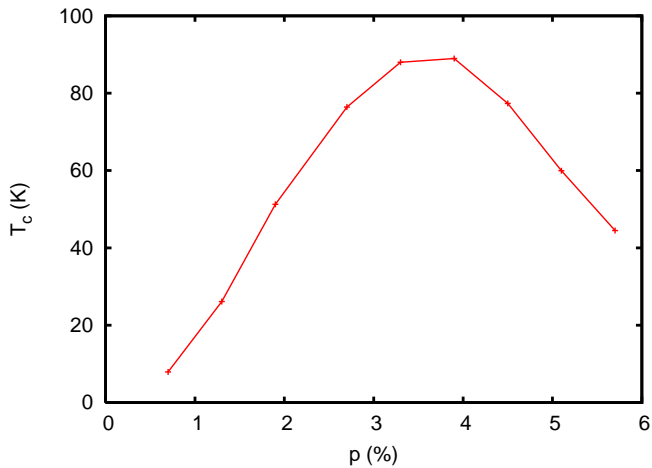}
\caption{Variation of $T_c$ with hole concentration $p$
for a $N=10^3$ system with $J=4$, $x=20\%$, and carrier bandwidth $W=12t=10$ eV,
averaged over 50 configurations.}
\end{figure}

The interplay of competing interactions and disorder is highlighted in Figure 6,
which shows a comparison of magnon DOS and $T_c$ for the ordered and disordered cases at 
the same dilution $x=1/8$.
Here the ordered case corresponds to a superlattice arrangement of impurities on alternate host lattice sites, 
studied earlier within a $k$-space sublattice-basis representation.\cite{diluted}
The low-energy part of the magnon spectrum, which corresponds to extended modes, 
is seen to be significantly softened in the disordered case, 
resulting in substantially reduced $T_c$, 
and simply reflects the reduced spin stiffness due to competing interactions. 
The high-energy magnon modes in the disordered case 
are due to magnon modes localized over strongly-coupled impurity spin clusters, as already discussed above.
The $T_c$ result for the ordered case is in close agreement with earlier $k$-space analysis
in terms of spin stiffness,\cite{diluted} 
providing additional check on the validity of finite-size $T_c$ calculations.

\begin{figure}
\hspace*{-3mm}
\includegraphics[width=90mm]{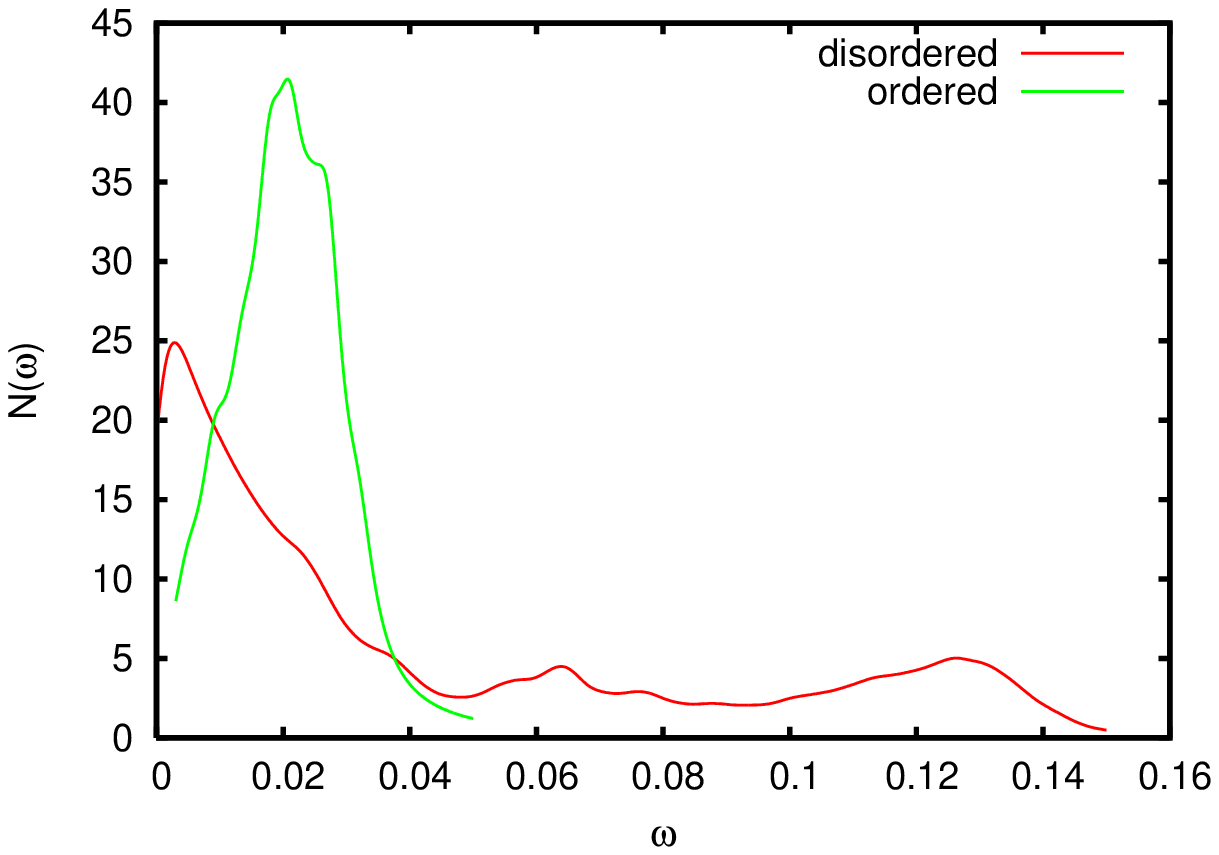}
\hspace*{-3mm}
\includegraphics[width=90mm]{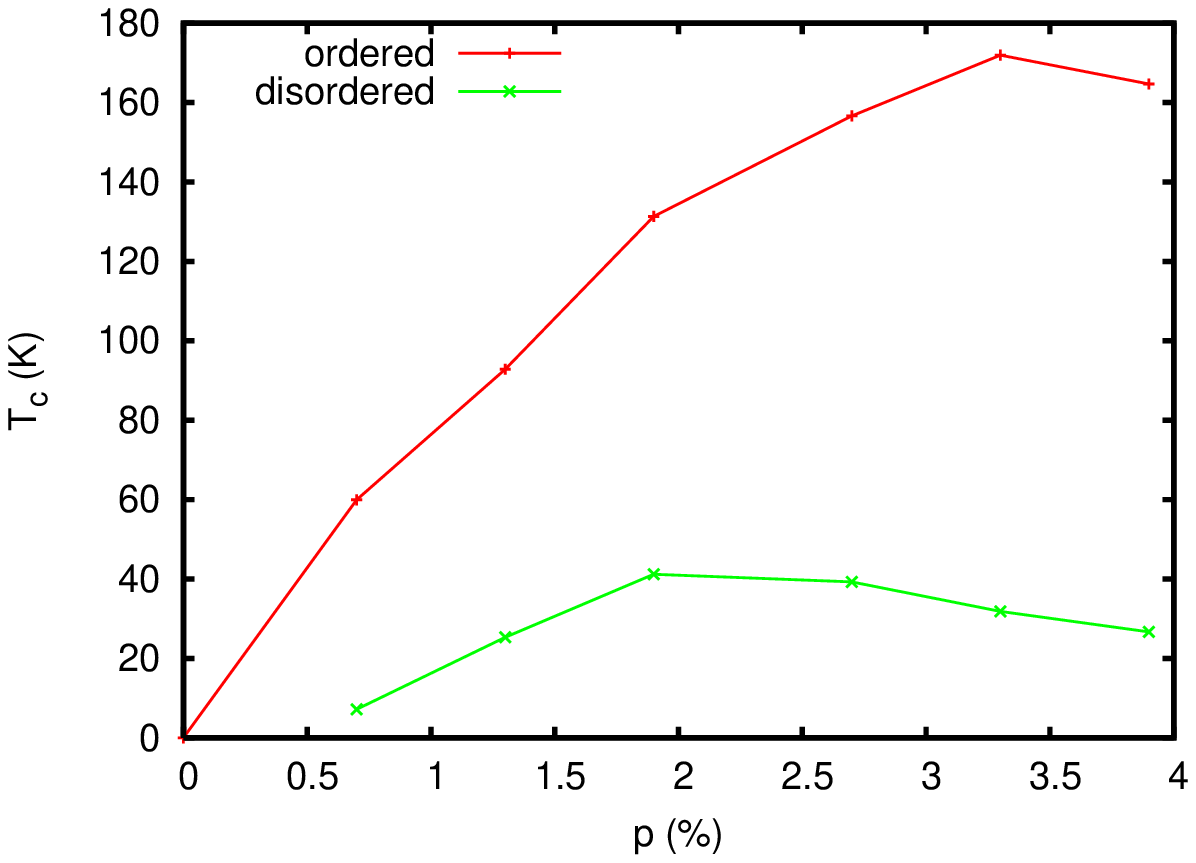}
\caption{Comparison of magnon DOS and $T_c$ for the ordered and disordered cases 
with exactly same dilution $x=1/8$, for a $N=10^3$ system with $J=4$. 
Here $p \approx 2.7\%$ in the upper panel and $W=10$ eV in the lower panel.}
\end{figure}

\section{Finite-temperature spin dynamics}
In section III we saw that dilution-induced disorder results in a strong enhancement 
in the density of low-energy magnons, with an appreciable fraction of localized modes corresponding
to weakly-coupled spins,
as well as formation of impurity-spin clusters supporting localized high-energy magnon modes.
In order to quantitatively investigate the effect of these magnon features
on the finite-temperature spin dynamics, 
in this section we evaluate the thermal reduction in magnetization due to magnon excitations. 

\begin{figure}
\hspace*{-3mm}
\includegraphics[width=90mm]{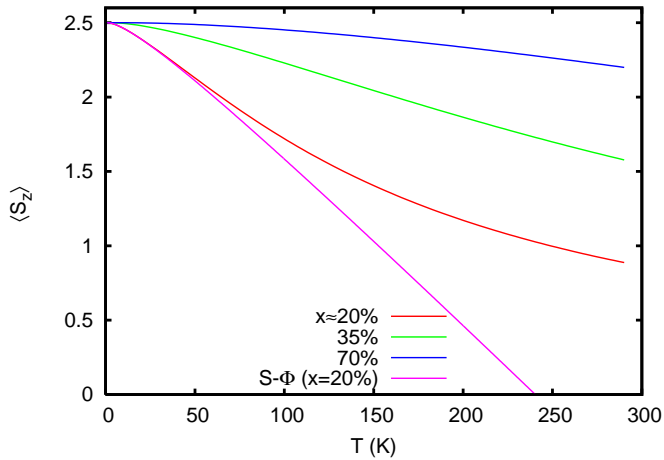}
\caption{Temperature dependence of averaged magnetization for different dilutions ($x\approx 20\%,35\%,70\%$)
and system sizes ($N=12^3,10^3,8^3$) while keeping the number of magnetic states fixed.
Here $J=4$, $W=10$eV, and $p/x =1/5$. Also shown is the conventional spin-wave-theory result (25).}
\end{figure}

We first discuss a locally self-consistent magnon renormalization scheme 
for the determination of local magnetizations, 
which is equivalent to the site-dependent Tyablikov decoupling procedure (local RPA) for a Heisenberg ferromagnet. 
We obtain the local magnetization from the Callen formula
\begin{equation}
\langle S_I ^z \rangle = \frac{
(S-\Phi_I)(1+\Phi_I)^{2S+1} + (S+1+\Phi_I)\Phi_I ^{2S+1} }
{(1+\Phi_I)^{2S+1} - \Phi_I ^{2S+1} }
\end{equation}
for a quantum spin-$S$ ferromagnet,\cite{callen}
where we have introduced site-dependent boson occupation numbers 
\begin{equation}
\Phi_I = \frac{1}{N_m} \sum_l \frac{|\phi_l ^I|^2} {e^{\beta \omega_l} - 1}
%(\langle S_z\rangle/S) 
\end{equation}
which explicitly involve the boson density $(\phi_{l}^I)^2$. 
A locally self-consistent magnon renormalization scheme (in terms of the fixed FKLM spin couplings 
${\cal J}_{IJ} \sim (J^2/4) [\chi^0]_{IJ}$) is then obtained
if the boson Hamiltonian matrix elements (10-13) are self-consistently renormalized 
\begin{eqnarray}
{\cal H}_{IJ} & = &
\sqrt{2\langle S_I^z\rangle} \left (\frac{J^2}{4}  [\chi^0]_{IJ} \right ) 
\sqrt{2\langle S_J^z \rangle} \nonumber \\
{\cal H}_{II} &=& \sum_{J\ne I} \left ( \frac{J^2}{4} [\chi^0]_{IJ} \right ) 2\langle S_J^z \rangle 
\end{eqnarray}
in terms of the local magnetization $\langle S_I^z \rangle$ instead of the MF values.
Together with ${\cal H}|\phi_l\rangle = \omega_l |\phi_l\rangle$,
the coupled equations (22-24) then self-consistently yield
the local magnetization $\langle S_I^z \rangle$ for all sites. 
In the translationally-symmetric case this yields the usual momentum-independent magnon-energy renormalization
$\omega_l = (\langle S^z \rangle /S) \omega_l^0$. 

\begin{figure}
\hspace*{-3mm}
\includegraphics[width=90mm]{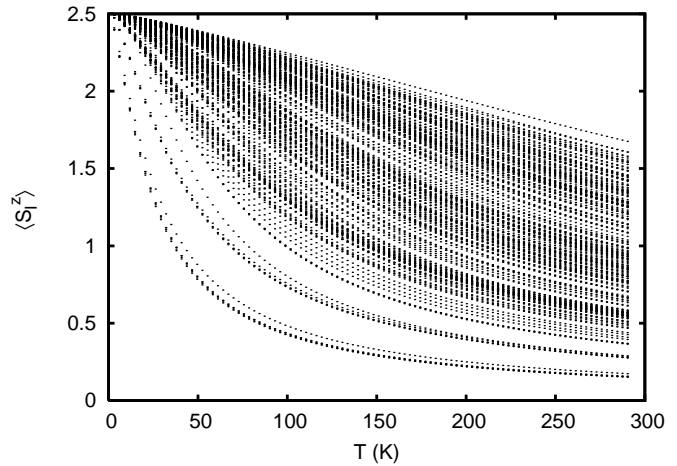}
\caption{Distribution of local magnetization, showing qualitatively different spin dynamics for weakly-
and strongly-coupled spins in a $N=10^3$ system with $J=4$, $x=20\%$, $p\approx 4\%$, and $W=10$eV.}
\end{figure}

In the low-temperature regime, thermal magnon renormalization is negligible
($\langle S_I ^z\rangle \rightarrow S$),
and the bare ($T=0$) magnon energies $\omega_l^0$ provide a good description of the spin dynamics.
We also have $\Phi_I << 1$,
and the magnetization equation (22) reduces to the conventional spin-wave-theory result 
for the site-averaged magnetization
\begin{equation}
\langle S_z \rangle = S- \frac{1}{N_m} \sum_I \Phi_I= S - \int d\omega \frac{N(\omega)}{e^{\beta \omega} - 1} \; ,
\end{equation}
in terms of the bare magnon density of states $N(\omega)$. 
On the other hand, in an intermediate-temperature regime
where $k_{\rm B}T \gg \omega_l^0$ for low-energy localized modes,
then $\Phi_I \sim k_{\rm B}T/\omega_l^0 >> 1$
and a nearly paramagnetic contribution $\langle S_I ^z \rangle \sim \omega_l^0 / k_{\rm B}T$ 
is obtained to the local magnetization, highlighting the qualitatively different spin dynamics of the
isolated, weakly-coupled spins. 

Figure 7 shows the temperature-dependence of magnetization evaluated from (22) and (23)
with the bare $(T=0)$ magnon energies and wavefunctions
and averaged over all impurity sites and several (20) configurations,
for different dilutions ($x\approx 20\%,35\%,70\%$) and fixed value of $p/x = 1/5$.
Different host system sizes $N=L^3$ with $L=12,10,8$ are taken so that 
the number of magnetic states per configuration remains same ($N_m=350$) in Eq. (23).
The strong enhancement in the thermal decay of magnetization with dilution
reflects the effect of the large enhancement in density of low-energy magnetic excitations 
on the finite-temperature spin dynamics. 
Also shown (for $x\approx 20\%$) is the conventional spin-wave-theory result (25), 
which asymptotically approaches the previous result in the low-temperature regime, as expected.

While the distinctly concave behaviour at higher temperature in the diluted case (Fig. 7, $x = 20\%$)
is in itself not conclusive evidence for nearly paramagnetic behaviour of weakly-coupled spins, 
as similar behaviour would be obtained even for a translationally-symmetric system
in a high-temperature ($k_{\rm B}T \gg \omega_l ^0$) regime,
the distribution of local magnetization $\langle S_I ^z \rangle$ for a single configuration,
clearly shows (Fig. 8) the qualitatively different spin dynamics for weakly- and strongly-coupled spins
{\em for the same magnon spectrum}, 
thus highlighting the role of the spatial character of magnon states. 
The wide range of behaviour for different spins, and therefore the different local environments
reflects the degree of complexity in a fully self-consistent magnon renormalization theory. 

\begin{figure}
\hspace*{-3mm}
\includegraphics[width=90mm]{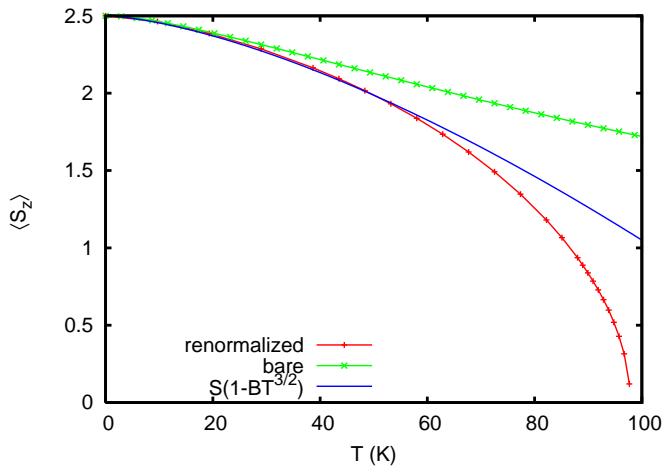}
\caption{Configuration-averaged magnetization within an approximate renormalization scheme 
involving the average boson occupation number,
for a $N=12^3$ system with $J=4$, $W=10$eV, $x \approx 20\%$, and $p\approx 4\%$,
along with a fit with the Bloch form $S(1-BT^{3/2})$, with $B=0.6\times10^{-3}\; {\rm K}^{-2/3}$.}
\end{figure}

%\begin{figure}
%\hspace*{-3mm}
%\includegraphics[width=90mm]{mt.eps}
%\caption{Temperature dependence of magnetization for a $N=8^3$ site system with different dilutions. Here $J=6$ and hole %doping $p \approx 4\%$.}
%$N=N_\uparrow =512$, and $N_\downarrow = 493$ (hole doping $p \approx 4\%$).}
%\end{figure}

We next include magnon renormalization within an approximate scheme,\cite{hilbert_nolting} 
suitable for the low-temperature regime where $\langle S_I ^z\rangle \sim S$ for all sites.
Here the average magnetization $\langle S_z\rangle$ is self-consistently calculated 
from an equation similar to (22)
in terms of the configuration- and site-averaged boson occupation number 
\begin{equation}
\Phi \equiv \langle \Phi_I \rangle = \left < \frac{1}{N_m} \sum_l \frac{1}
{e^{\beta (\langle S_z\rangle/S) \omega_l^0} - 1} \right >_c
\end{equation}
where the bare ($T=0$) magnon energies $\omega_l^0$ are uniformly renormalized by the factor $\langle S_z\rangle/S$,
as in the standard Tyablikov theory.
Figure 9 shows the temperature dependence of average magnetization 
obtained for several ($N_c=20$) configurations of a $N=12^3$ system with $J=4$, $x\approx 20\%$, and $p\approx 4\%$.
As expected, in the low-temperature regime where $\langle S_z\rangle \rightarrow S$ and $\Phi_I << 1$,
the result approaches the previous (unrenormalized) result.
Interestingly, the low-temperature behaviour of magnetization 
fits well with the Bloch form $S(1-BT^{3/2})$, with $B=0.6\times10^{-3}\; {\rm K}^{-2/3}$, 
which is of same order of magnitude (1.4$\times10^{-3}\; {\rm K}^{-2/3}$) 
as obtained in squid magnetization measurements of 
an annealed 50nm sample of $\rm Ga_{1-x} Mn_x As $ with $6\%$ Mn concentration.\cite{sperl}. 

\begin{figure}
\hspace*{-3mm}
\includegraphics[width=90mm]{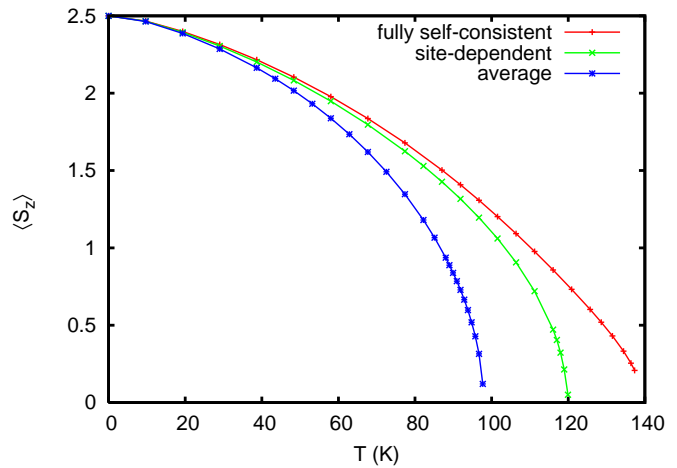}
\caption{Substantial enhancement in $T_c$ due to spin clustering 
when the site-dependence of local magnon occupation number and magnetization is included,
for a $N=10^3$ system (single configuration), with $J=4$, $W=10$eV, $x = 20\%$, and $p\approx 4\%$.}
\end{figure}

Dealing only with the average magnon occupation number,
the above approximation does not incorporate the spatial segregation of magnon modes,
particularly of the localization of high-energy modes over strongly-coupled impurity-spin clusters.
Whereas, strong local correlations and high disordering temperature 
in these spin clusters should generally enhance long-range ordering and $T_c$,
and therefore a quantitative analysis of this spin clustering is of interest.
Localization of high-energy magnon modes over impurity clusters implies that 
cluster spins have relatively smaller participation in the low-energy modes, 
resulting in smaller magnon occupation number $\Phi_I$ and higher local magnetization $\langle S_I^z \rangle$ 
on cluster sites at low temperature. 
Fig. 10 shows the substantial enhancement in $T_c$ resulting from including the site-dependence of 
$\Phi_I$ and $\langle S_I^z \rangle$, evaluated self-consistently from (22) and (23), 
but with a uniform magnon-energy renormalization 
$\omega_l = (\langle S^z \rangle /S) \omega_l^0$ in (23)
in terms of the average magnetization $\langle S^z \rangle = (1/N_m) \sum_I \langle S_I^z \rangle$.
Also shown is the result of the fully self-consistent magnon renormalization scheme (22-24),
where the renormalized magnon energies $\omega_l$ and wave functions $\phi_l$ are obtained  
by diagonalizing the renormalized magnon Hamiltonian (24) which is updated at every step.

\begin{figure}
\hspace*{-3mm}
\includegraphics[width=90mm]{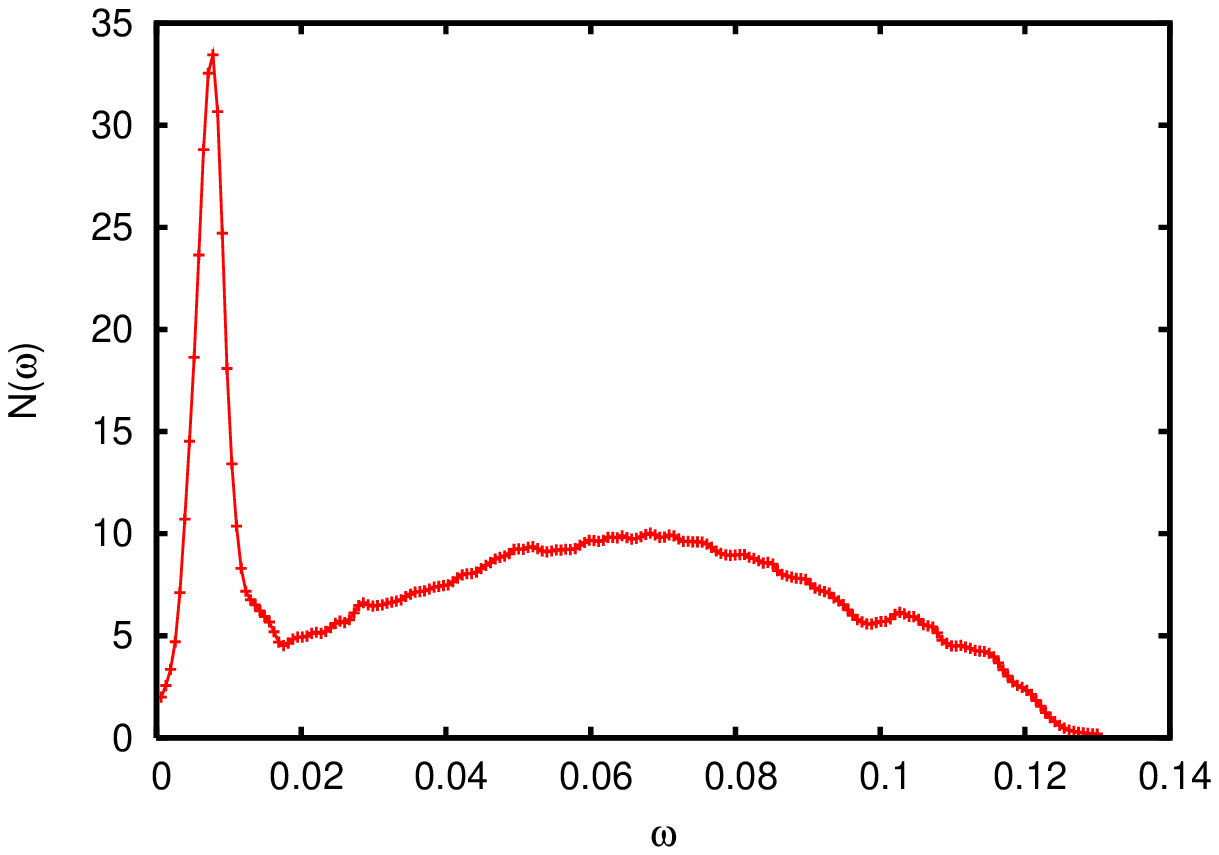}
%\end{figure}
%\begin{figure}
\hspace*{-3mm}
\includegraphics[width=90mm]{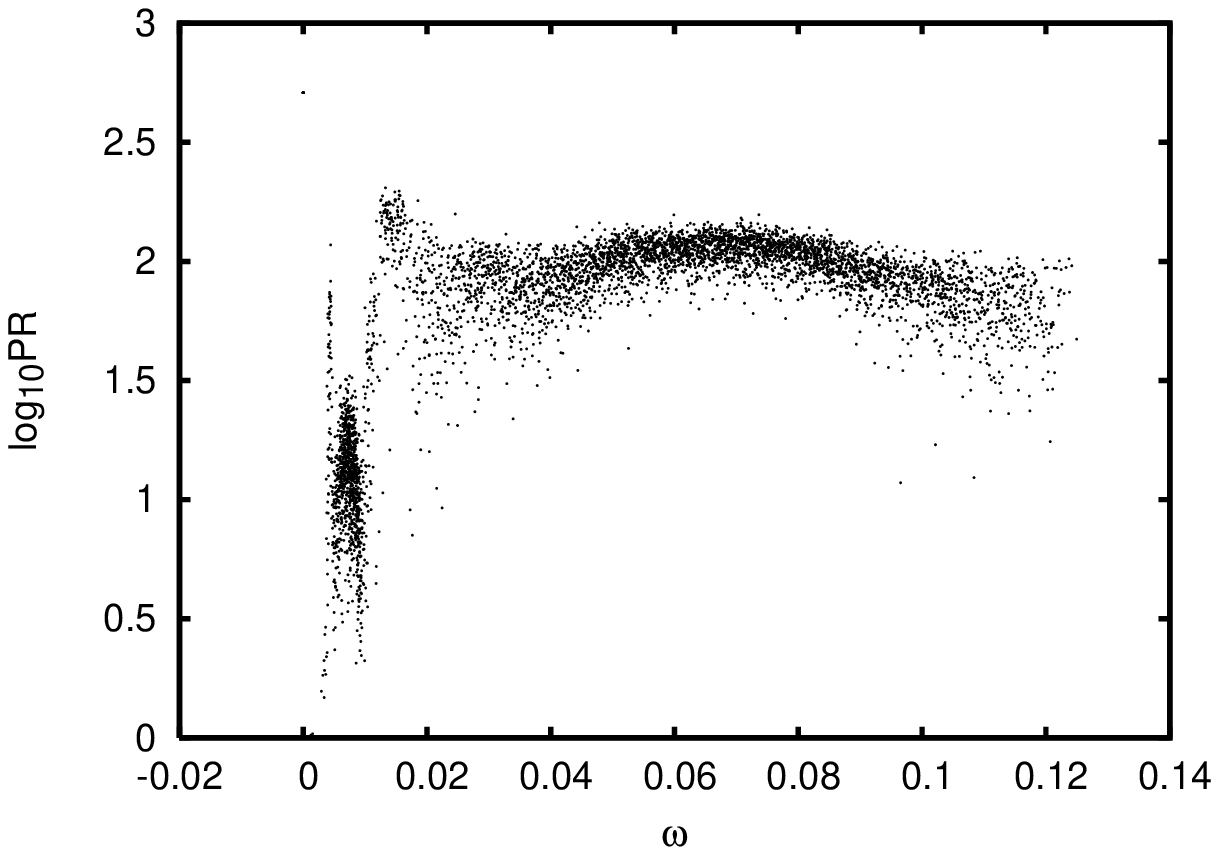}
\caption{Magnon density of states and PR for the FKLM with a small fraction $(x'=N'/N \approx 20\%)$ 
of weakly-coupled spins $(t'=0.4)$ in a $N=8^3$ system with $J=4$ and $p \approx 11\%$.}
\end{figure}

\section{FKLM with weakly-coupled spins}
Even at $x\approx 20\%$ impurity concentration, 
the dominant low-temperature magnetization behaviour (Fig. 9) in the diluted FKLM  
is determined by the majority low-energy extended modes,
although a finite fraction of weakly-coupled spins is also clearly present (Fig. 8). 
In order to highlight the contribution of the low-energy localized modes,
in this section we consider a FKLM with a small fraction of weakly-coupled spins 
(resembling nearly paramagnetic impurities), which has some resemblance to the impurity-band model,\cite{dis2}
and evaluate the lattice-averaged magnetization within the locally self-consistent magnon renormalization scheme
discussed in section IV. We therefore consider a FKLM 
\begin{equation}
H = \sum_{I,\delta,\sigma} t_I a_{I,\sigma}^\dagger a_{I+\delta,\sigma} 
- \frac{J}{2} \sum_I{\bf S}_I.{\mbox{\boldmath $\sigma$}}_I
\end{equation}
where the nearest-neighbour hopping terms $t_I = t'<t$ for a small fraction ($x'=N'/N$) of randomly chosen sites,
representing relatively isolated and weakly-coupled spins, 
within an impurity-band representation.\cite{dis2} 

\begin{figure}
\hspace*{-3mm}
\includegraphics[width=90mm]{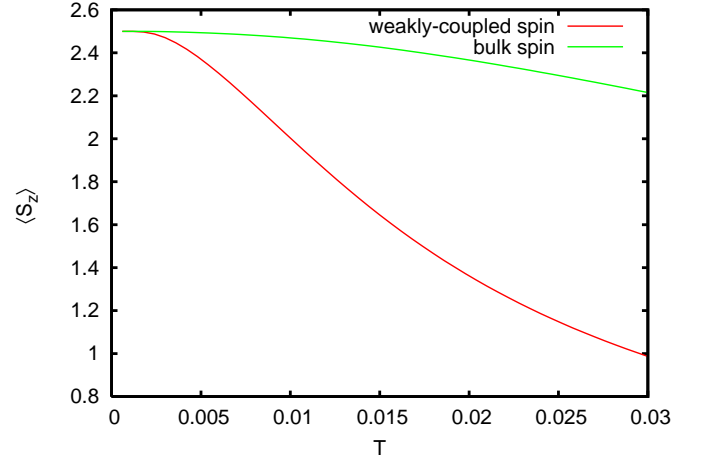}
\caption{Comparison of temperature dependence of local magnetizations $\langle S_I ^z\rangle$ 
of a weakly-coupled spin $(t'=0.4)$ and a bulk spin in a $N = 8^3$ system with $J=4$ and $p \approx 11\%$.}
\end{figure}

\begin{figure}
\hspace*{-3mm}
\includegraphics[width=90mm]{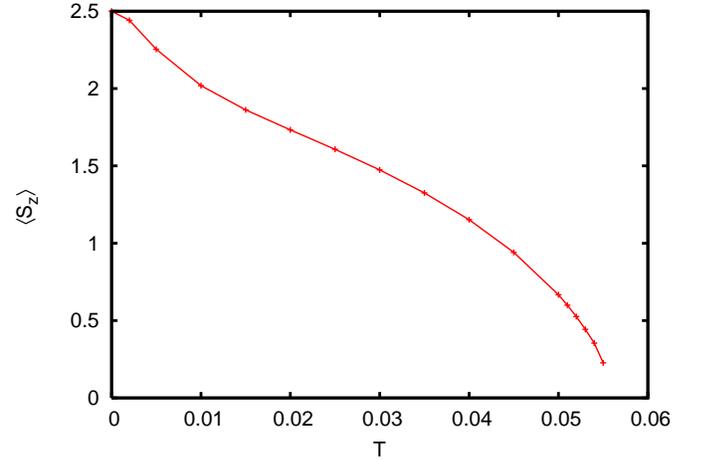}
\caption{Lattice-averaged magnetization obtained within the locally self-consistent magnon renormalization scheme 
for the FKLM with $N'=150$ weakly-coupled spins $(t'=0.3)$ in a $N=8^3$ spin system
with $J=4$ and $p \approx 11\%$.}
\end{figure}

The magnon DOS and IPR (Fig. 11) clearly show low-energy localized modes corresponding to weakly-coupled spins,
qualitatively similar to the diluted KLM case with randomly placed impurity spins,
but with much sharper differentiation between extended (bulk) and localized (impurity) magnon states. 
The disorder here is relatively weaker and apart from the small fraction of low-energy magnon modes, 
remaining modes are essentially unaffected.
For a single weakly-coupled spin in the lattice, 
Fig. 12 shows the qualitatively different (unrenormalized) spin dynamics for the weakly-coupled and bulk spins
obtained from Eqs. (22) and (23) with the bare magnon energies $\omega_l^0$.  
Results of the locally self-consistent magnon renormalization scheme (22-24) are shown in Fig. 13,
which highlights both the dominant nearly paramagnetic spin dynamics of weakly-coupled spins in the 
low-temperature regime, and the dominant spin dynamics of bulk spins near $T_c$.
 
\section{FKLM with potential disorder}
Competing interactions between the randomly separated impurities due to positional disorder 
is an important feature of carrier-induced ferromagnetism in the diluted FKLM.
In order to highlight the interplay of positional disorder and competing interactions, and 
distinguish the role of purely potential disorder, 
we consider the concentrated FKLM with potential disorder
\begin{equation}
H = t \sum_{I,\delta,\sigma} a_{I,\sigma}^\dagger a_{I+\delta,\sigma} 
+ \sum_{I,\sigma} \epsilon_I a_{I,\sigma}^\dagger a_{I,\sigma} 
- \frac{J}{2} \sum_I{\bf S}_I.{\mbox{\boldmath $\sigma$}}_I
\end{equation}
where the random on-site energies are distributed in the range $-W/2 < \epsilon_I < W/2$. 
with a rectangular probability distribution of width $W$.
Magnon energies are again obtained following the general method discussed in section II, 
by diagonalizing the magnon Hamiltonian (10) involving the effective spin couplings $J^2 \chi_{IJ}$ 
in terms of the particle-hole propagator $\chi_{IJ}$.
As there is no dilution, the number of magnetic states $N_m$ is equal to the number of sites $N$.
Without disorder, magnons in the concentrated FKLM have been studied earlier 
in the context of heavy fermion materials,\cite{sig} ferromagnetic metals Gd, Tb, Dy, 
doped EuX\cite{donath} and manganites.\cite{furu,wang,yunoki,vogt} 

\begin{figure}
\hspace*{-3mm}
\includegraphics[width=90mm]{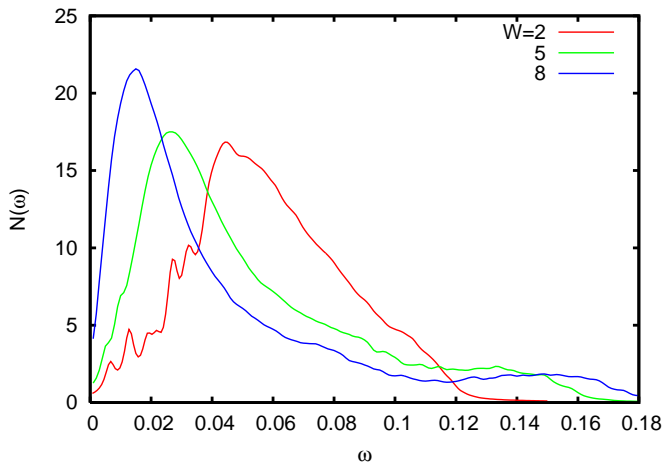}
\caption{Magnon density of states for a $N=8^3$ system with different potential disorder strength $W$.  
Here $J=4$ and hole doping $p \approx 11\%$.}
\end{figure}

We find that purely potential disorder has a relatively milder effect on the magnon spectrum (Fig. 14), 
with a gradual shift in the magnon density of states peak to lower energy, 
indicating gradual decrease of $T_c$ with disorder.
While the magnon spectrum is substantially softened,
there are no negative-energy magnon modes even for disorder strength comparable to the bandwidth,
and the collinear ferromagnetic state remains robust.
The narrowing of the magnon energy distribution in Fig. 14 is a typical feature of the disordered ferromagnet 
in which the low-energy extended modes are relatively less affected 
by the small averaged-out disorder on long length scales.
By comparison, the interplay of competing interactions and disorder in the diluted case 
results in much stronger magnon softening (Fig. 2),
and even with a comparable potential disorder $\sim JS/2=5$ for $J=4$,
increasing dilution results in strong magnon softening and emergence of finite fraction of
negative-energy modes indicating instability of the collinear ferromagnetic state.

For the concentrated FKLM without disorder, the spin-stiffness expression consists of two parts --- 
a positive delocalization-energy term of order $t$ favouring ferromagnetic ordering 
and a negative exchange-energy term of order $t^2/JS$ favouring antiferromagnetic ordering.  
The magnon softening seen in Fig. 14 due to disorder can be readily understood in terms of a disorder-induced 
enhancement of the average exchange-energy term $\sim \langle t^2/(JS + \epsilon_I -\epsilon_J) \rangle $,
as also obtained for the disordered Hubbard model.\cite{singh+vollhardt} 
Essentially, enhancement in the exchange term when the potential barrier is reduced 
(negative $\epsilon_I -\epsilon_J$) more than compensates for its reduction when the barrier is enhanced  
(positive $\epsilon_I -\epsilon_J$), resulting in enhanced average exchange energy. 

\section{Conclusions}
The exact treatment of disorder in our finite-size calculations
allowed for a quantitative investigation of the interplay of disorder and competing interactions 
in the carrier-induced ferromagnetic state of the diluted ferromagnetic Kondo lattice model.
Quantitative measures of the competition between impurity spin couplings,
stability of the ferromagnetic state, and the magnetic transition temperature 
could be obtained from our real-space analysis of magnon excitations,
with the estimated $T_c$ in good agreement with experimental results for $\rm Ga_{1-x}Mn_x As$
for corresponding impurity concentration, hole bandwidth, and compensation.
Finite-temperature spin dynamics was studied within a locally self-consistent magnon renormalization scheme,
equivalent to a site-dependent Tyablikov decoupling (local RPA),
to investigate the role of magnon localization over isolated, weakly-coupled spins and over strongly-coupled spin clusters.

The strong magnon softening observed with dilution reflects
increasingly effective competition between carrier-induced spin couplings,
especially at lower hole doping where the couplings are longer-ranged.
Indeed, the instability of the collinear ferromagnetic state at high dilution, 
as signalled by appearance of negative-energy magnon modes,
indicates strong competition between ferromagnetic and antiferromagnetic spin couplings.
Highlighting the importance of compensation in DMS systems, 
the strong optimization of $T_c$ with hole doping at $p/x << 1$ 
stems from the competition between increasing overall magnitude and increasing rapidity
of oscillation of the carrier-induced spin couplings. 
Comparison of the ordered and disordered cases for exactly same dilution, 
with significant magnon softening and lowering of $T_c$ in the disordered case,
explicitly demonstrates the interplay of disorder and competing interactions in DMS systems.
The relatively milder magnon softening and robustness of ferromagnetic state obtained for
the concentrated FKLM with purely potential disorder serves to highlight 
the greater degree of frustration due to positional disorder in diluted systems. 

The large enhancement in density of low-energy magnetic excitations due to competing interactions and disorder 
is responsible for the strong thermal decay of magnetization at high dilution. 
While the distribution of local magnetization clearly exhibits 
spin-dynamics behaviour of both weakly- and strongly-coupled spins,
the low-temperature behaviour of lattice-averaged magnetization was found to be 
dominated by low-energy extended magnon states, 
and fitted well with the Bloch form $M_0(1-BT^{3/2})$, with $B$ of the same order of magnitude as obtained
in recent squid magnetization measurements on $\rm Ga_{1-x}Mn_x As$ samples,\cite{sperl}

The enhancement in ferromagnetism due to strong local correlations in strongly-coupled spin clusters
is captured in the locally self-consistent magnon renormalization scheme,
as reflected in the substantial enhancement in $T_c$ obtained
on including the site dependence of magnon occupation numbers.
Furthermore, applied to the FKLM with a small fraction of weakly-coupled spins, 
the scheme highlighted both the dominant nearly-paramagnetic spin dynamics of weakly-coupled spins 
in the low-temperature regime, and the dominant spin dynamics of bulk spins near $T_c$.
Thus a wide range of spin-dynamics behaviour involving different local environments
can be studied practically within the self-consistent magnon renormalization scheme.

Finally, we mention the outstanding and interesting issue of fermion-sector renormalization,
in particular of the carrier-induced spin couplings 
due to quantum and thermal corrections to the bare particle-hole propagator $[\chi^0(\omega)]$. 
In the VCA-type approaches, the effective impurity field seen by fermions is taken to vanish
as $T\rightarrow T_c$, yielding RKKY-type spin couplings.
However, the large separation between moment-melting and moment-disordering temperatures
implies presence of appreciable local moments even near $T_c$,
so that fermions should continue to see impurity fields 
due to the slowly-fluctuating, locally-ordered impurity moments. 
Indeed, the splitting of fermion bands near $T_c$ 
even for rather moderate couplings ($J/W \sim 0.2$),
obtained within dynamical self-energy studies of the concentrated FKLM,\cite{nolting}
is precisely due to presence of the slowly-fluctuating impurity fields.
It will be of interest to examine both quantum and thermal corrections to spin couplings due to
self-energy and vertex corrections in $[\chi^0(\omega)]$
within a spin-rotationally invariant scheme which preserves the Goldstone mode.
Studied recently for the ferromagnetic state of the Hubbard model,\cite{vertex} 
the net quantum correction can be essentially understood in terms of an exchange-energy correction 
due to fermion spectral-weight transfer.
Preliminary calculation of spectral-weight transfer for the FKLM indicates a suppresion by the factor $1/S$,
suggesting relatively smaller quantum corrections for large spin quantum number $S$.

\ \\
\ \\
\section*{Acknowledgement}
It is a pleasure to thank M. Sperl for communicating to us results prior to publication
and G. X. Tang, J. Kienert, and S. Henning for helpful discussions.
A. Singh acknowledges support from the Alexander von Humboldt Foundation.


\begin{thebibliography}{99}

\bibitem{ohno2}H. Ohno, A. Shen, F. Matsukara, A. Oiwa, A. Endo, S. Katsumoto, 
and Y. Iye, Appl. Phys. Lett. {\bf 69}, 363 (1996).

\bibitem{matsu}F. Matsukara, H. Ohno, A. Shen, and Y. Sugawara, Phys. Rev. B 
{\bf 57}, R2037 (1998).

\bibitem{ohno3}H. Ohno and F. Matsukara, Solid State Commun. {\bf 117}, 179 (2003).

\bibitem{ku}K. C. Ku, S. J. Potashnik, R. F. Wang, S. H. Chun, P. Schiffer, N. Samarth, 
M. J. Seong, A. Mascarenhas, E. Johnston-Halperin, R. C. Myers, 
A. C. Gossard, and D. D. Awschalom, Appl. Phys. Lett. {\bf 82}, 2302 (2003). 

\bibitem{edmonds}K. W. Edmonds, P. Boguslawski, K. Y. Wang, R. P. Campion, 
S. N. Novikov, N. R. S. Farley, B. L. Gallagher, C. T. Foxon, M. Sawicki, T. Dietl, 
M. B. Nardelli, and J. Bernholc, Phys. Rev. Lett. {\bf 92}, 037201 (2004).

\bibitem{mf1}
T. Dietl, A. Haury, and Y. M. d'Aubigne,
Phys. Rev. B {\bf 55}, R3347 (1997);
T. Dietl, H. Ohno, F. Matsukara, J. Cibert, and D. Ferrand, 
Science, {\bf 287}, 1019 (2000);
T. Dietl, F. Matsukara, and H. Ohno,
Phys. Rev. B {\bf 66}, 033203 (2002). 

\bibitem{mf2}
M. Takahashi, Phys. Rev. B {\bf 56}, 7389 (1997).

\bibitem{mf3}
T. Jungworth, W. A. Atkinson, B. H. Lee, and A. H. MacDonald, 
Phys. Rev. B {\bf 59}, 9818 (1999);
B. H. Lee, T. Jungworth, and A. H. MacDonald,
{\em ibid} {\bf 61}, 15606 (2000).

\bibitem{sw1}
J. K\"{o}nig, H. H. Lin, and A. H. MacDonald, 
Phys. Rev. Lett. {\bf 84}, 5628 (2000);
J. K\"{o}nig, T. Jungworth, and A. H. MacDonald, 
Phys. Rev. B {\bf 64}, 184423 (2001);
J. Schliemann, J. K\"{o}nig, and A. H. MacDonald,
Phys. Rev. B {\bf 64}, 165201 (2001).

\bibitem{dis2}
M. Berciu and R. N. Bhatt,
Phys. Rev. Lett. {\bf 87}, 107203 (2001);
Phys. Rev. B {\bf 66}, 085207 (2002).

%\section{dmft1}
\bibitem{dmft1}
A. Chattopadhyay, S. Das Sarma, and A. J. Millis,
Phys. Rev. Lett. {\bf 87}, 227202 (2001).

\bibitem{sw2}
G. Bouzerar and T. P. Pareek, 
Phys. Rev. B {\bf 65}, 153203 (2002).

\bibitem{dis1}
J. Schliemann and A. H. MacDonald,
Phys. Rev. Lett. {\bf 88}, 137201 (2002).

\bibitem{dis3}
A. L. Chudnovskiy and D. Pfannkuche, 
Phys. Rev. B {\bf 65}, 165216 (2002).

\bibitem{dis6}
C. Timm, F. Sch\"{a}fer, and F. von Oppen,
Phys. Rev. Lett. {\bf 89}, 137201 (2002);
C. Timm and F. von Oppen, J. Supercond. {\bf 16} 23 (2003);
cond-mat/0209055 (2002).

\bibitem{anomalous}
M. P. Kennett, M. Berciu and R. N. Bhatt,
Phys. Rev. B {\bf 65}, 115308 (2002); {\em ibid} {\bf 66}, 045207 (2002).

\bibitem{mc3}
G. Alvarez, M. Mayr, and E. Dagotto,
Phys. Rev. Lett. {\bf 89}, 277202 (2002);
G. Alvarez and E. Dagotto,
Phys. Rev. B {\bf 68}, 045202 (2003).

\bibitem{dassarma}
S. Das Sarma, E. H. Hwang, and A. Kaminski, Phys. Rev. B {\bf 67}, 155201 (2003);
D. J. Priour, Jr., E. H. Hwang, and S. Das Sarma, Phys. Rev. Lett. {\bf 92}, 117201 (2004).

\bibitem{dms}A. Singh, A. Datta, S. K. Das, and V. A. Singh,
Phys. Rev. B {\bf 68}, 235208 (2003). 

\bibitem{squid}
A. Singh, cond-mat/0307009 (2003).

%reviews 
\bibitem{bhatt}R. N. Bhatt, M. Berciu, M. P. Kennet, and X. Wan,
Jour. of Superconductivity INM {\bf 15}, 71 (2002).

\bibitem{timm}C. Timm, J. Phys.: Condens. Matter {\bf 15}, R1865 (2003).

\bibitem{sperl}
M. Sperl, J. Sadowski, R. Gareev, W. Wegscheider, D. Weiss, and G. Bayreuther,
to be published. 

\bibitem{sawicki}
M. Sawicki, K.-Y. Wang, K. W. Edmonds, R. P. Campion, C. R. Staddon, N. R. S. Farley,
C. T. Foxon, E. Papis, E. Kami\'{n}ska, A. Piotrowska, T. Dietl, and B. L. Gallagher,
Phys. Rev. B {\bf 71}, 121302(R) (2005); 
M. Sawicki, J. Magn. Magn. Mater. {\bf 300}, 1 (2006).

\bibitem{kipferl}
W. Kipferl, M. Sperl, T. Hagler, R. Meier, and G. Bayreuther,
J. Appl. Phys. {\bf 97}, 10B313 (2005);
M. Sperl, W. Kipferl, M. Dumm, and G. Bayreuther,
{\em ibid} {\bf 99}, 08J703 (2006).

\bibitem{notredame}
D. M Wang, Y. H. Ren, R. Merlin, K. Dziatkowski, X. Liu, J. K. Furdyna, and M. Grimsditch,
to appear in ICPS proceedings (2006).

\bibitem{sig}M. Sigrist and K. Ueda, and H. Tsunetsugu, Phys. Rev. B {\bf 46}, 
175 (1992); M. Sigrist, H. Tsunetsugu, K. Ueda, and T. M. Rice, 
Phys. Rev. B {\bf 46}, 13838 (1992).

\bibitem{donath}M. Donath, P. A. Dowben, and W. Nolting, eds., {\it Magnetism and 
Electronic Correlations in Local-Moment Systems: Rare-Earth Elements and Compounds} 
(World Scientfic, Singapore, 1998).

%\bibitem{Nol1}W. Nolting, W. M\"uller and C. Santos, cond-mat/0304338 (2003). 

\bibitem{furu}N. Furukawa, J. Phys. Soc. Jpn. {\bf 65}, 1174 (1996).

\bibitem{wang}X. Wang, Phys. Rev. B {\bf 57}, 7427 (1998).

\bibitem{yunoki}S. Yunoki, J. Hu, A. L. Malvezzi, A Moreo, N. Furukawa, and E. Dagotto, 
Phys. Rev. Lett. {\bf 80}, 845 (1998); E. Dagotto, S. Yunoki, A. L. Malvezzi, A. Moreo, 
J. Hu, S. Capponi, D. Poilblanc, and N. Furukawa, Phys. Rev. B {\bf 58}, 6414 (1998).

\bibitem{vogt}M. Vogt, C. Santos, and W. Nolting, Phys. Stat. Sol. (b)
{\bf 223}, 679 (2001).

\bibitem{nol2}
W. Nolting, T. Hickel, A. Ramakanth, G. G. Reddy, and M. Lipowczan, 
Phys. Rev. B {\bf 70}, 075207 (2004).

\bibitem{diluted}
S. K. Das and A. Singh, cond-mat/0506523 (2005).

\bibitem{hilbert_nolting}
S. Hilbert and W. Nolting, Phys. Rev. B {\bf 70}, 165203 (2004);
Phys. Rev. B {\bf 71}, 113204 (2005).

\bibitem{bouzerar}
G. Bouzerar, T. Ziman, and J. Kudrnovsk\'{y},
Europhys. Lett. {\bf 69}, 812 (2005); G. Bouzerar, T. Ziman, and J. Kudrnovsk\'{y},
Appl. Phys. Lett. {\bf 85}, 4941 (2004); 
R. Bouzerar, G. Bouzerar, and T. Ziman, Phys. Bev. B {\bf 73}, 024411 (2006). 

\bibitem{vertex}
A. Singh, cond-mat/0512648. 

\bibitem{dag06}
F. Popescu, Y. Yildirim, G. Alvarez, A. Moreo, and E. Dagotto,
cond-mat/0601593.

\bibitem{sharma_06}
A. Sharma and W. Nolting, to be published. 

\bibitem{callen}
H. B. Callen, Phys. Rev. {\bf 130}, 890 (1963). 

\bibitem{singh+vollhardt}
A. Singh, M. Ulmke, and D. Vollhardt, Phys. Rev. B {\bf 58}, 8683 (1998).

\bibitem{nolting}
W. Nolting, S. Rex, and S. Mathi Jaya, J. Phys. Cond. Matt. {\bf 9}, 1301 (1997).

\end{thebibliography}
\end{document}